\documentclass{jpp}
\usepackage{graphicx}
\usepackage{epstopdf, epsfig}

 \usepackage{amssymb}
 \usepackage{bigints}
 \usepackage{amsmath}
 \usepackage{empheq}
 \usepackage{framed}
 \usepackage{color}
 \usepackage{enumerate}
 \usepackage{ulem}
 \usepackage{bm}
 \usepackage{appendix}
 \usepackage{multicol}
 \numberwithin{equation}{section}
 \usepackage{upgreek}
 \usepackage{textgreek}
\usepackage{scalerel}
\usepackage{stackengine,wasysym}
\usepackage{stmaryrd} 
\usepackage{xfrac}    

\usepackage{subfigure}

\newcommand{\comment}[1]{}

\newcommand{\be}{\begin{equation}}
\newcommand{\ee}{\end{equation}}
\newcommand{\ba}{\[\begin{aligned}}
\newcommand{\ea}{\end{aligned}\]}
\newcommand{\bea}{\begin{eqnarray}}
\newcommand{\eea}{\end{eqnarray}}
\newcommand{\beann}{\begin{eqnarray*}}
\newcommand{\eeann}{\end{eqnarray*}}
\newcommand{\bs}{\begin{split}}
\newcommand{\es}{\end{split}}

\newcommand*{\cB}{\mathcal{B}}

\newcommand*{\cE}{\mathcal{E}}
\newcommand*{\cF}{\mathcal{F}}
\newcommand*{\cG}{\mathcal{G}}

\newcommand*{\ep}{\epsilon}

\newcommand*{\B}{\bm{B}}

\newcommand*{\J}{\bm{J}}

\newcommand*{\vpl}{v_{||}}

\newcommand*{\dpl}{\nabla_{||}}

\newcommand*{\Jpl}{J_{||}}

\newcommand*{\dl}{\bm{\nabla}}
\newcommand*{\del}{\partial}
\newcommand*{\BD}{\bm{B}\cdot\bm{\nabla}}

\newcommand*{\lbr}{\left(}
\newcommand*{\rbr}{\right)}

\newcommand{\iotabar}{\mbox{$\iota\!\!$-}}

\newcommand*{\psibar}{\overline{\psi}}
\newcommand*{\alphabar}{\overline{\alpha}}

\newcommand*\reallywidetilde[1]{\ThisStyle{%
  \setbox0=\hbox{$\SavedStyle#1$}%
  \stackengine{-.1\LMpt}{$\SavedStyle#1$}{%
    \stretchto{\scaleto{\SavedStyle\mkern.2mu\sim}{.5467\wd0}}{.7\ht0}%
  }{O}{c}{F}{T}{S}%
}}

\def\test#1{$%
  \reallywidetilde{#1}\,
  \scriptstyle\reallywidetilde{#1}\,
  \scriptscriptstyle\reallywidetilde{#1}
$\par}
\makeatletter
\newsavebox{\@brx}
\newcommand{\llangle}[1][]{\savebox{\@brx}{\(\m@th{#1\langle}\)}%
  \mathopen{\copy\@brx\mkern2mu\kern-0.9\wd\@brx\usebox{\@brx}}}
\newcommand{\rrangle}[1][]{\savebox{\@brx}{\(\m@th{#1\rangle}\)}%
  \mathclose{\copy\@brx\mkern2mu\kern-0.9\wd\@brx\usebox{\@brx}}}
\makeatother

\shorttitle{Particles near X-point} 

\title{ Charged particle dynamics near an X-point of a non-symmetric magnetic field with closed field lines}

\shortauthor{E. Elbarmi, W. Sengupta, H. Weitzner}
\author{Elena Elbarmi\aff{1,2}, Wrick Sengupta\aff{1}, Harold Weitzner\aff{1}}

\affiliation{\aff{1}Courant Institute of Mathematical Sciences, New York University, New York, New York 10012, USA
\aff{2} Friends Seminary, 22 E 16th St, New York, NY 10003, USA
}

\begin{document}

\maketitle

\begin{abstract}

Understanding particle drifts in a non-symmetric magnetic field is of primary interest in designing optimized stellarators to minimize the neoclassical radial loss of particles. Quasisymmetry and omnigeneity, two distinct properties proposed to ensure radial localization of collisionless trapped particles in stellarators, have been explored almost exclusively for magnetic fields that generate nested flux surfaces. In this work, we extend these concepts to the case where all the field lines are closed. We then study charged particle dynamics in the exact non-symmetric vacuum magnetic field with closed field lines, obtained recently by Weitzner and Sengupta (arXiv:1909.01890), which possesses X-points. The magnetic field can be used to construct magnetohydrodynamic equilibrium in the limit of vanishing plasma pressure. Expanding in the amplitude of the non-symmetric fields, we explicitly evaluate the omnigeneity and quasisymmetry constraints. We show that the magnetic field is omnigeneous in the sense that the drift surfaces coincide with the pressure surfaces. However, it is not quasisymmetric according to the standard definitions.
 
\end{abstract}

\section{Introduction \label{sec:intro}}

Intrinsic steady-state and disruption-free operations and lack of a density limit are some of the significant advantages of stellarators over tokamaks. However, the fully three-dimensional nature of stellarator geometries introduces several mathematical, physical, and engineering challenges \citep{helander2012comparison, helander2014theory}.  
The existence of magnetohydrodynamic (MHD) equilibrium or vacuum magnetic fields in a non-symmetric torus with nested flux surfaces or exclusively closed field lines is not well understood. Even in systems with continuous symmetry where nested flux surfaces are guaranteed to exist, it is easily shown that symmetry breaking by small perturbations can lead to the formation of magnetic islands and stochastic regions near the rational surfaces. The pressure and rotational transform profiles cannot be, therefore, arbitrarily chosen to avoid the breakup of rational surfaces \citep{grad1967toroidal,newcomb1959magnetic,hudsonKrauss20173D_cont_B}. The difficulties in obtaining exact solutions or even perturbative analytic expressions for three-dimensional non-symmetric vacuum magnetic fields with surfaces have been pointed out \citep{cary1982vacuum,freidberg_idealMHD, sengupta_weitzner2019_lowshear_vacuum}. 

Broadly there are three kinds of vacuum or equilibrium magnetic field lines relevant to toroidal confinement, namely, ergodic field lines on nested flux surfaces, ergodic volume filling field lines, and closed field lines. Ergodic field lines lying on flux surfaces only occur in magnetic field line systems that are integrable \citep{caryLittlejohn1983Hamiltonian_B} when the rotational transform is irrational. In the absence of any continuous symmetry, the magnetic field line system is, in general chaotic, and the field lines are ergodic and volume filling. Closed magnetic fields are found in systems with continuous symmetries when the rotational transform is rational \citep{newcomb1959magnetic}, as well as in non-symmetric systems with discrete symmetries \citep{grad1971plasma,lortz1970existenz}. The only rigorous example of an ideal MHD equilibrium without any continuous symmetry is given by  \cite{lortz1970existenz}. He proved that in a system with reflection symmetry, a non-symmetric three-dimensional MHD equilibrium with a smooth pressure profile can be constructed iteratively starting with a vacuum field with closed field lines and zero shear. A significant drawback of this result is that the rotational transform for such an equilibrium is zero. Weitzner and Sengupta (2019) derived exact analytical expressions for vacuum magnetic fields with closed field lines in a flat toroidal shell ( a topological torus) and extended Lortz's result to include all rational rotational transform.

Even if non-symmetric equilibrium or vacuum fields with approximately nested surfaces can be constructed, radial confinement of particles trapped in these fields can not be guaranteed. It is well known that particle orbits in perfectly axisymmetric magnetic fields are periodic and closed in the radial direction. In non-axisymmetric magnetic fields with flux surfaces, circulating particle orbits are radially periodic. In contrast, trapped particles generally have trajectories that gradually drift out from the confined region outward to open field lines over a period of a few bounce times. The neoclassical particle and energy fluxes associated with these unconfined orbits can be significant for un-optimized stellarators \citep{mynick2006transport}. Various properties of the magnetic field such as quasisymmetry \citep{nuhrenberg1988quasihelical,boozer1995quasi}, quasi-axisymmetry \citep{nuhrenberg1994quasiaxi,okamura2001quasiaxi}, omnigeneity \citep{hall1975three,caryshashrinaPRL1997helical} and quasi-isodynamicity  \citep{gori_Lotz_varenna1996,nuhrenberg2010isodynamic} have been proposed to minimize the radial drifts of particles. In a stellarator with either of these properties, neoclassical transport is reduced \citep{beidler2001improved,hirsch2008majorW7AS,HSX_canik2007exp_neoclassical, w7x_klinger2016performance} and particle confinement is enhanced \citep{subbotin2006integrated,skovoroda20053dimproved}.  

Collisionless guiding center theory for single-particle motion shows that the secular radial drift of particles is absent in magnetic field systems with nested flux surfaces when the second adiabatic invariant $\Jpl=\oint \vpl dl $ is independent of the magnetic field line label on a given flux surface. Omnigeneity is defined as configurations where surfaces of constant $\Jpl$, known as the drift surfaces, are also flux surfaces \citep{caryshasharina1997omnigenity}. Quasisymmetry (QS) can be shown to be a particular case of omnigeneity \citep{landremancatto2012omnigenity}. Whereas omnigeneity is a nonlocal property that needs the information of the bounce points of a trapped particle, QS is local and requires more stringently that the strength of the magnetic field has a continuous symmetry. Through a local expansion near a minimum of the magnetic field, strength \citep{sengupta_weitzner2018} shows that it is easier to satisfy the omnigeneity condition than the quasisymmetry requirement.

Using near-axis asymptotic expansions \citep{garrenBoozer1991existence,landreman_Sengupta2018direct,landreman_Sengupta_Plunck2019direct,landreman_Sengupta2019_2nd_order}, it has been shown that the QS constraint makes the MHD system overdetermined, and in general, such magnetic fields might not exist in a given volume. Lack of QS in a volume intrinsically deteriorates the confinement of $\alpha$ particles, and careful numerical optimization is required \citep{bader2019stellarator}.
While the search for finding exact QS in a volume is still ongoing \citep{burby_Kallinikos_MACKAY2019_Math_QS}, it has been shown that QS can be exactly satisfied on one surface \citep{plunk2018quasiaxi_vacuum, garrenBoozer1991existence}. The freedom in choosing this surface can be cleverly exploited to minimize fast particle loss \citep{henneberg2019fast_particle}. Interestingly, numerical results show that the best choice of the surface for the optimization of quasi-axisymmetry is somewhat midway between the magnetic axis and the edge of the plasma.

Geometry also leads to significant differences in the tokamak and stellarator divertor programs \citep{konig2002divertor,feng2011comparison, helander2012comparison}. Tokamaks usually have poloidal-field divertors which respect the toroidal symmetry, whereas modern stellarators either have helical or island divertors with multiple X-points. Both of these are three-dimensional and much more complicated than poloidal-field divertors. The field-line pitch in an island divertor, unlike the tokamak divertor, arises from the magnetic shear and not from the rotational transform. In low-shear stellarators, the island divertors, therefore, have much larger connection lengths than tokamak divertors. Larger connection length implies that the cross-field transport is more important in the island divertor than in a tokamak divertor. The competition between cross-field and parallel transport leads to complex flow and plasma profiles \citep{HSX_akerson2016three}. Cross-field transport is sensitive to the magnetic field geometry, and a proper understanding requires a fully three-dimensional analysis \citep{feng2006physics}.

In this work, we study charged particle dynamics in a vacuum magnetic field with closed field lines \citep{weitzner_sengupta2019exact} and  X-points. We restrict ourselves to a flat toroidal shell (a topological torus), which addresses all the complexities arising from the doubly periodic nature of the toroidal domain without the complexities associated with the toroidal curvature. The model magnetic field has all closed field lines with zero shear and reflection symmetries. Experimental results from Wendelstein VII-A/AS \citep{hirsch2008majorW7AS,brakel2002energytransp_rational_iota_W7AS,brakel1997W7AS_EB_shear} and numerical results \citep{wobig1987localized_pert_w7A,andreevaW7Xvacuum} support the idea that optimum confinement is usually found close to certain low-order rational surfaces. Linear \citep{grad1973magnetofluid,nelson_spies_1974_suff_MHD_closed,hameiri1980_MHD_local_stability,nelson_spies_1974_suff_MHD_closed} and nonlinear  MHD stability analysis \citep{spies1974nonlinear_stability_closed} show that a closed field system can be stable even if neighboring low-shear systems are unstable. Furthermore, the necessary stability conditions obtained assuming shear might not be relevant for low-shear systems \citep{grad1973magnetofluid,spies1979low_report}.

In studying QS systems, the existence of MHD equilibrium with nested flux surfaces or integrability of field line flow is generally assumed \citep{helander2014theory}, although it is not necessary \citep{burby_Kallinikos_MACKAY2019_Math_QS}. Since rational surfaces where the field lines close on themselves are measure zero compared to irrational surfaces in integrable magnetic field systems, it is tacitly assumed in most QS and omnigeneity analysis that the field lines are ergodic on the flux surfaces. In this work, we formulate and analyze the closed field line versions of these concepts without the assumption of any continuous symmetry. In particular, we explore the relation between drift surfaces and pressure surfaces in closed field line systems. In such systems, pressure surfaces are labeled by surfaces of constant $q=\oint dl/B$ \citep{grad1971plasma}, where $B$ is the magnetic field strength, and the integration is carried out along a closed field line.
In contrast to MHD equilibrium with irrational rotational transform, pressure surfaces in closed field line systems are not necessarily functions of only one Clebsch variable. Hence, a proper distinction must be made between radial confinement strategies for systems with rational and irrational rotational transforms. We construct explicit examples to compare and contrast the omnigeneity and QS definitions for these two systems. In particular, we show that even though the system is not omnigeneous or QS according to the standard definitions for irrational surfaces, the drift surfaces for all particles can coincide with pressure surfaces ($q$ surfaces), making the system omnigeneous. This has implications for the particle dynamics near the X-point as well. Through an explicit calculation of the lowest order drift surfaces, we show that the constant $q$ surfaces play a critical role in determining the particle drifts near the X-points if the magnetic shear is negligible. Since QS and omnigeneity, in general, can not be exactly satisfied, it does not matter if the drift surfaces ( $\oint dl \vpl$) and the $q$ surfaces ($\oint dl/B$ ) surfaces do not coincide to higher orders \citep{benDaniel1965nonexistence,grad1967toroidal}. Therefore, the lowest order asymptotic analysis can provide valuable insight into the collisionless cross-field drifts of particles in systems with very long connection lengths.

The outline of the paper is as follows. In section \ref{sec:omni_QS_closed}, we shall discuss the concepts of omnigeneity and quasisymmetry in the magnetic field system with closed field lines. In section \ref{sec:drift_surfaces_closed}, we shall discuss the model magnetic field with closed field lines and evaluate the omnigeneity and quasisymmetry constraints. We shall compare the lowest order drift surfaces with the constant $q$ surfaces and the Clebsch flux variable. For the model magnetic, the lowest order drift surfaces are shown to coincide with the constant $q$ surfaces. Finally, we shall discuss the implications of our results in section \ref{sec:discussion}.

\section{Omnigeneity and quasisymmetry for magnetic field systems with closed field lines}
\label{sec:omni_QS_closed}

Ideal MHD equilibrium with magnetic fields that close on themselves and with fields that do not close but lie on flux surfaces have many intrinsic differences \cite{grad1971plasma}. For ergodic field lines with flux surfaces, pressure surfaces coincide with the flux surfaces. For closed field lines, pressure surfaces coincide with surfaces of constant $q=\oint dl/B$ and $\oint B dl $ \citep{grad1971plasma,spies1974nonlinear_stability_closed}, where $B$ is the strength of the magnetic field and the integration is performed along the closed field line. Since $q$ is not necessarily a flux function, pressure and flux surfaces do not have to coincide.
It is not surprising, therefore, that quasisymmetry and omnigeneity would also show characteristic differences for closed and ergodic field lines with flux surfaces. In particular, the requirement
that the contour for the global maximum of the field strength of omnigeneous and quasisymmetric fields must be a straight line in Boozer coordinates \cite{caryshasharina1997omnigenity,caryshashrinaPRL1997helical, landremancatto2012omnigenity,plunk_Landreman_2019near_axis3}, is valid only for ergodic field lines that lie on irrational flux surfaces. This condition, together with Cary-Shasharina mapping \citep{caryshashrinaPRL1997helical,caryshasharina1997omnigenity, plunk_Landreman_2019near_axis3} show that ergodic magnetic fields with flux surfaces that are omnigeneous but not quasisymmetric must be non-analytic. Given these differences, we present a brief description of vacuum and MHD equilibrium with closed field lines before presenting the definitions of quasisymmetry and omnigeneity in such closed field line systems.

\subsection{MHD and vacuum fields with closed lines}
Ideal MHD equilibrium with scalar pressure, $p$, is governed by the equations
\begin{align}
\dl\cdot \B =0, \quad \J= \dl \times \B, \quad  \J\times\B=\dl p.
\label{ideal_MHD_eqb}
\end{align}
As a consequence we have
\begin{align}
    \J\cdot \dl p=0, \quad \B\cdot \dl p=0.
    \label{MHD_eqb}
\end{align}
In the following we shall consider magnetic fields of only two types: ergodic field lines on nested flux surfaces and closed field lines. In these two cases we can use a single-valued function $q$ to label pressure surfaces, i.e. $p=p(q)$ \citep{grad1971plasma}. Since $\J$ is divergence-free and orthogonal to $\dl p$, following Grad and Weitzner \citep{grad1971plasma, weitzner2014ideal}, we use a Clebsch representation for the current
\begin{align}
     \J = \dl \zeta \times \dl q = \dl \times (\zeta \dl q).
     \label{Jform}
\end{align}

Ideal MHD force balance equation then implies
\begin{align}
\BD \zeta = p'(q).
\label{MDE_zeta}
\end{align}

Since $\J=\dl \times \B$, equation \eqref{Jform} shows that we can represent magnetic field $\B$ in the form
\begin{align}
    \B = \dl \Phi + \zeta \dl q.
    \label{Bcov}
\end{align}

With the help of the current potential $\zeta$ and magnetic scalar potential $\Phi$, we can reformulate the ideal MHD equilibrium conditions 
\eqref{ideal_MHD_eqb} as 
\begin{align}
    \BD q=0,\quad \BD \zeta =p'(q), \quad  \B = \dl \Phi + \zeta \dl q, \quad \dl \cdot \B =0
    \label{idealMHD_phizeta}
\end{align}
The primary advantage of the above formulation is that it brings to light the mixed hyperbolic and elliptic nature of the MHD equilibrium \citep{grad1971plasma,weitzner2014ideal}, that leads to mathematical and computational difficulties  \citep{Garabedian2012computational} in solving \eqref{ideal_MHD_eqb} in non-symmetric toroidal domains.

We shall now examine \eqref{idealMHD_phizeta} in a toroidal domain where $q$ denotes a ``radial" coordinate, $\varphi$ denotes a ``toroidal" angle and $\theta$ denotes a ``poloidal" angle. Physical quantities must be $2\pi$ periodic in the two angles. The functions $\Phi$ and $\zeta$ and their gradients are in general multi-valued in such a toroidal domain. Since $\B$ and $\J$ must be single-valued functions, the periods of $\Phi$ and $\zeta$ must satisfy additional relations. From \eqref{Jform} we find that the multi-valued part of $\dl\zeta$ must be in the $\dl q$ direction since $\J$ is single-valued. Equation \eqref{Bcov} then shows that the periods of $\Phi$ must be functions of $q$. Furthermore, it follows from the condition $\BD p(q)=0$ that $\Phi$ and $\zeta$ satisfy
\begin{align}
 \zeta =-\frac{\dl q \cdot \dl \Phi}{|\dl q|^2}, \quad \text{such that} \quad \B =\frac{\dl q \times (\dl\Phi \times \dl q)}{|\dl q|^2}.
 \label{zetaEqn}
\end{align}
Therefore, $\Phi$ and $\zeta$ must be of the form
\begin{align}
    \Phi=F(q) \theta +G(q)\varphi +\tilde{\Phi}(q,\theta,\varphi), \quad
    \zeta=-(F'(q) \theta  +G'(q) \varphi  ) +\tilde{\zeta}(q,\theta,\varphi),
    \label{PhiZetafom}
\end{align}
where, $\tilde{\Phi}$ and $\tilde{\zeta}$ are periodic functions of the angles $\theta$ and $\varphi$.

Finally, to satisfy $\dl \cdot \B=0$ we may use the Clebsch potentials $\psi,\alpha$ such that
\begin{align}
    \B=\dl \psi \times \dl \alpha.
    \label{Clebsch_form_B}
\end{align}
Here we assume that $\psi$ is single-valued while $\alpha$ is not. The choice of $\psi=q$ can be made but is not necessary. The multi-valued function $\alpha$ must have periods that are functions of $\psi$ so that $\B$ is single-valued. Grad considered the MHD formulation \eqref{idealMHD_phizeta} for closed field line systems with zero rotational transform. \cite{weitzner2014ideal} extended the formalism to include arbitrary rotational transform by equating the form \eqref{Bcov} for the magnetic field to the Clebsch form \eqref{Clebsch_form_B} with the choice of $\psi=q$, and using \eqref{zetaEqn} to eliminate $\zeta$ in \eqref{idealMHD_phizeta}. In our notation the following is Weitzner's formulation of ideal MHD
\begin{subequations}
\begin{align}
\dl q \times \dl \alpha = \frac{\dl q \times (\dl\Phi \times \dl q)}{|\dl q|^2}, \quad \dl q\times \dl \alpha \cdot \dl \lbr \frac{\dl q\cdot \dl \Phi}{|\dl q|^2}\rbr=-p'(q).
\end{align}
 \label{weitzner_MHD_formalism}
\end{subequations}
The first equation represents the generalized Cauchy-Riemann equations that couple the angular derivatives of $\Phi$ and $\alpha$ on a constant $q$ surface. The second equation is the generalized Grad-Shafranov equation for non-symmetric MHD equilibrium \citep{weitzner2014ideal,weitzner2016expansions}.

The above description applies equally well to both closed field lines and ergodic field lines with flux surfaces. For an ergodic field line on a constant $\psi$ surface, $p$ being single-valued can be a function of only $\psi$. We can choose $\psi$ to be the toroidal flux. The function $q$, in this case, is a function of $\psi$ alone. Since $\B$ is single-valued, $\alpha$ must be of the form \citep{weitzner2014ideal,kruskal_Kulsrud_1958equilibrium}
\begin{align}
    \alpha= f(\psi)\: \theta + g(\psi) \:\varphi + \tilde{\alpha}(\psi,\theta,\varphi),
    \label{alpha_def}
\end{align}
where $\tilde{\alpha}$ is a periodic function. In flux coordinates, $\tilde{\alpha}=0$ and the rotational transform is given by the twist
\begin{align}
    \iotabar(\psi)=\frac{\BD \theta}{\BD \varphi}=-\frac{g(\psi)}{f(\psi)}.
\end{align}

For field lines that closes on themselves after $n$ poloidal ($\theta$) and $m$ toroidal ($\varphi$) circuit, $\iotabar = n/m$ i.e. a rational number \cite{newcomb1959magnetic}. In the expression for the field-line label \eqref{alpha_def}, we can absorb any $\psi$ dependence from $f,g$ by redefining $\psi$ since the ratio $f/g$ is a constant. Hence $\alpha$ is of the form
\begin{align}
    \alpha= m \theta - n \varphi +\tilde{\alpha},
\end{align}
which shows that $\dl \alpha$ is single-valued. The function $q$ can be any arbitrary function such that $\BD q=0$. In particular it can depend on both $\psi$ and $\alpha$ i.e. $q=q(\psi,\alpha)$. 

We now choose the Clebsch variable $q$ in a closed field line system, motivated by a study of the system \eqref{idealMHD_phizeta}. 
Following \cite{grad1971plasma}, we use the notation $[K]$ to denote the jump in a multi-valued function $K$ after a complete circuit along the closed field line. Integrating \eqref{MDE_zeta} along the closed field line we find that the jump in $\zeta$ is given by
\begin{align}
    [\zeta]= \oint \bm{dl}\cdot \dl \zeta= p'(q) \oint \frac{dl}{B}.
\end{align}
Similarly, from \eqref{Bcov} and \eqref{zetaEqn} we obtain the magnetic field circulation
\begin{align}
    \oint \B\cdot\bm{dl} = \oint \bm{dl}\cdot \dl \Phi = [\Phi]\quad \text{with}\quad [\Phi]' = -[\zeta].
\end{align}
Equation \eqref{PhiZetafom} shows that
\begin{align}
    [\Phi]= 2\pi (n F(q)+m G(q)), \quad   [\Phi]'=2\pi(n F'(q)+m G'(q))=-[\zeta].
    \label{PhiZetabox}
\end{align}
Therefore, identifying $q$ with $\oint dl/B$ we get
\begin{align}
    [\zeta]=q p'(q), \quad  [\Phi]=\oint \B\cdot\bm{dl} = -\int q p'(q) dq +\text{constant},
    \label{Circulation}
\end{align}
which proves that level sets of pressure $p$, magnetic field circulation $\oint Bdl=[\Phi]$, and $q=\oint dl/B$ coincide.

For vacuum fields, we can simply substitute $\zeta = p'=0$ in the above equations. Equation \eqref{PhiZetafom} with $\zeta=0$ imply that for vacuum fields $F$ and $G$ must be constants. From \eqref{Circulation}, we find that the magnetic field circulation for vacuum fields is also constant. 

\subsection{Drifts, quasisymmetry and omnigeneity for  field lines with flux surfaces }
For field lines that do not close but lie on flux surfaces, the leading order gyroaveraged radial drift of a particle of mass $m$ and charge $e$ is given by \citep{helander2014theory}
\begin{align}
    \bm{v_d}\cdot \dl \psi =\cF\:\vpl\dpl \lbr \frac{\vpl}{\Omega} \rbr ,
    \label{radial_psi_drift}
\end{align}
where $\Omega= e B/m$ is the gyrofrequency, $\vpl=\sigma\sqrt{2(\cE-\mu B)}$ is the parallel speed of the particle with total energy $\cE$, magnetic moment $\mu$, and $\sigma=\text{sign}(\vpl)$. The factor $\cF$ is a geometric quantity given by
\begin{align}
    \cF=\frac{\B\cdot \dl \psi \times \dl B}{\BD B}.
    \label{cFdef}
\end{align}
Bounce averaging the radial drift we obtain the secular radial excursion
\begin{align}
    \Delta \psi= \oint \frac{dl}{\vpl} \bm{v_d}\cdot \dl \psi = \oint dl \cF\: \dpl \lbr \frac{\vpl}{\Omega} \rbr.
    \label{psi_excursion}
\end{align}
For trapped particles the endpoints of the integral corresponds to the bounce points, whereas for circulating particles that sample the entire torus, the line integral $\oint dl$ approaches a flux surface average \citep{grad1967toroidal,taylor_hastie,helander2014theory}. Since the bounce integral for a circulating particle do not depend on the position, the bounce average can be replaced by a flux-surface average \citep{grad1967toroidal, hazeltine_meiss2003plasma_confinement_book}. It can then be shown \citep{helander2014theory} that the radial excursion for a circulating particle is identically zero. This argument does not hold for trapped particles because the bounce points are $\alpha$ dependent.

It is well known \citep{gardner1959adiabatic,helander2014theory,caryshasharina1997omnigenity,sengupta_weitzner2018} that the secular radial drift is related to the derivative of the second adiabatic invariant $\Jpl=\oint \vpl dl$ with respect to the field line label $\alpha$ through
\begin{align}
\Delta \psi = \frac{1}{(e/m)}\frac{\del \Jpl}{\del \alpha}.
    \label{d_alpha_Jpl}
\end{align}
To avoid the secular radial excursion of particles $\Jpl$ therefore must be independent of $\alpha$, i.e. \begin{align}
    \Jpl=\Jpl(\psi).
    \label{omnigeneity}
\end{align}
Magnetic fields with this property are called omnigeneous \citep{caryshashrinaPRL1997helical,helander2014theory}. The drift surfaces (constant $\Jpl$) coincide with the pressure surfaces (constant $\psi$) in omnigeneous magnetic fields. Since the evaluation of $\Jpl$ for trapped particles requires information on the bounce points, the omnigeneity condition \eqref{omnigeneity} is nonlocal. For circulating particles on an irrational flux surface that ergodically samples the whole torus, there is no intrinsic dependence on the initial or the final positions. Hence the omnigeneity condition \eqref{omnigeneity} is satisfied. However, trapped particles, especially deeply trapped particles, only sample a small part of the torus, and therefore they do not necessarily satisfy the omnigeneity condition.

There is a local condition called quasisymmetry \citep{landremancatto2012omnigenity,Lima_Paul_Wright_2019introduction, burby_Kallinikos_MACKAY2019_Math_QS} that is stricter than omnigeneity but guarantees that $\Delta \psi=0$ is satisfied identically. 
There are various equivalent definitions of QS \citep{helander2014theory, Landreman_Simons_Summer_lecture,burby_Kallinikos_MACKAY2019_Math_QS}. We restrict ourselves strictly to fields that are either vacuum or satisfy MHD equations. The two most commonly used coordinate-free definitions of QS for ergodic magnetic fields with flux label $\psi$ are
\begin{align}
    \BD \lbr \frac{\B\cdot \dl \psi \times \dl B}{\BD B}\rbr=0, \quad \text{and} \quad \dl \psi \times \dl B \cdot \dl \lbr \BD B\rbr=0.
    \label{QS1_psi}
\end{align}
The above equations can be integrated to obtain the following slightly more recognizable form of QS conditions
\begin{align}
  \cF=  \frac{\B\cdot \dl \psi \times \dl B}{\BD B}= \cF(\psi), \quad \text{and} \quad \BD B= f(\psi,B).
  \label{QS2_psi}
\end{align}
Since $\dpl \cF=0$, the condition \eqref{psi_excursion} is automatically satisfied. 

\subsection{Drifts, quasisymmetry and omnigeneity for closed field lines }
We now discuss the subtleties involved in trying to extend the previous expressions for the QS and omnigeneity conditions in a closed field line system. Mainly there are two issues. Firstly, in a closed field line system, field lines with different values of $\alpha$ can have different pressures since $p=p(\psi,\alpha)$. Mathematically, for a given $\B$ and its two Clebsch potentials (characteristics) $(\psi,\alpha)$, we can always form new coordinates $(\psibar(\psi,\alpha),\alphabar(\psi,\alpha))$ such that $\B=\dl \psibar \times \dl \alphabar$. Once a choice is made, say for $\psibar$, we can solve for $\alphabar$ by equating the Poisson bracket of $(\psibar,\alphabar)$ with respect to $(\psi,\alpha)$ to one i.e. 
\begin{align}
\{\psibar,\alphabar\}_{(\psi,\alpha)}=\frac{\del \psibar}{\del \psi}\frac{\del \alphabar}{\del \alpha}- \frac{\del \psibar}{\del \alpha}\frac{\del \alphabar}{\del \psi}=1
    \label{PB_psialpha_bar}
\end{align}

Secondly, for closed field line systems, the $\Jpl$ invariant for both circulating particles and trapped particles can be dependent on $\alpha$, and hence both species can contribute to the lack of omnigeneity. The standard argument of replacing the bounce integral by a surface integral fails for circulating particles in a closed field line system since they do not sample the entire torus. This has been pointed out earlier by \cite{grad1967toroidal,taylor_hastie} and recently in \citep{burby_Kallinikos_MACKAY2019_Math_QS}. The problem persists even if the magnetic shear is low if the rotation transform is close to a rational surface, such that a circulating particle takes a very long time to sample the torus. The surface average definition of $\Jpl$ is inaccurate \citep{grad1967toroidal}, so is the formula of $\Jpl=\oint dl \vpl$ \cite{taylor_hastie}. 

Given the degeneracy in choosing the Clebsch variables, we now make a particular choice of $\psibar$, based on physical grounds. In ergodic systems, localization of trapped particles in omnigeneous systems is sought near the flux-surfaces, which are also pressure surfaces. We can extend the same idea to closed field line systems by seeking radial localization near pressure surfaces. Since pressure is a function of $q=\oint dl/B$, we can choose $\psibar=q(\psi,\alpha)$. The corresponding $\alphabar$, obtained from solving \eqref{PB_psialpha_bar}, is given by
\begin{align}
    \alphabar= \int\limits_{q=\text{const}}\frac{d\alpha}{(\del q / \del\psi)} \:.
\end{align}

We now propose to modify the QS and omnigeneity conditions for closed field lines based on the physical motivation that pressure surfaces (equivalently $q=\oint dl/B$ surfaces) be used to measure the radial drift and the radial excursion. The relevant formulas are
\begin{align}
    \bm{v_d}\cdot \dl q = \cG \vpl \dpl \lbr \frac{\vpl}{\Omega }\rbr,\: \: \Delta q = \oint dl\: \cG\: \dpl \lbr \frac{\vpl}{\Omega} \rbr \quad \text{with} \quad \cG = \frac{\B \times \dl q \cdot \dl B}{\BD B}.
\end{align}
Omnigeneity will now be defined by the condition that the second adiabatic invariant for both trapped and untrapped species of particles be function of only $q$ and independent of $\alphabar$, 
\begin{align}
\Jpl=\oint \vpl dl =\Jpl(q), \quad \frac{\del \Jpl}{\del \alphabar}=0. 
    \label{omni_closed}
\end{align}
For trapped particles, the line integral is carried out between bounce points, and for passing particles, the integral is over the whole closed field line. 

Physically, this implies that drift surfaces (constant $\Jpl$) and the pressure surfaces (constant $\oint dl/B$) coincide. Interestingly, if such a closed field line configuration exists, then it automatically satisfies the low plasma-beta necessary and sufficient interchange stability conditions in the fluid as well as kinetic stability \citep{benDaniel1965nonexistence,taylor1963_stability_cusp,taylor_hastie,taylor_hastie_1968stability}. Such configurations might be very difficult to achieve exactly without a continuous symmetry \citep{benDaniel1965nonexistence,grad1967toroidal}. However, this is not a significant concern since omnigeneity and QS are themselves seldom exactly obtained.

The QS conditions are obtained by replacing $\psi$ with $q$ in \eqref{QS2_psi}. The proofs of these are identical to the ones presented in \cite{helander2014theory}.
\begin{align}
\frac{\B\cdot \dl q \times \dl B}{\BD B}= \cG(q), \quad \text{and} \quad \BD B= f(q,B).
  \label{QS2_q}
\end{align}
Since $\dpl \cG=0$, $\Delta q=0$. Therefore, closed field line QS, according to \eqref{QS2_q}, implies closed field line omnigeneity \eqref{omni_closed}. However, the converse need not be true, just like that in magnetic fields with flux surfaces.

Finally, we note that another characteristic property of QS, namely, the periodicity of $B$ \citep{helander2014theory}, is preserved as well in a closed field line system. We have given a brief proof in the appendix \ref{appendix_QS_periods}. We shall now analyze drift orbits in a model vacuum magnetic field system and examine whether the drift motions can be localized near a $q$ surface.
 
\section{Drift orbits in a nonsymmetric magnetic field with closed field lines} \label{sec:drift_surfaces_closed}
\cite{weitzner_sengupta2019exact} recently obtained an exact doubly-periodic vacuum magnetic field by solving the fully three-dimensional Laplace's equation in a topological torus. Assuming a discrete symmetry, they showed that all the magnetic field lines are closed. Starting with this vacuum magnetic field and using Lortz's iteration, one can obtain ideal MHD equilibrium in a topological with a smooth pressure profile and any rational iota. Recent analytical and numerical results \citep{kim2019elimination} show that smooth pressure profiles can indeed be supported in a topological torus with self-consistent boundary conditions. Linearizing about an equilibrium with zero magnetic shear, they show that the singularity on a rational surface is removable.

We shall first briefly describe the model vacuum magnetic field, its characteristics $(\psi,\alpha)$, and the function $q=\oint dl/B$. Our model is also relevant to non-symmetric MHD equilibrium, since the the closed field line equilibrium pressure must be a function of the vacuum $q$ to the lowest order in the plasma-beta. We shall then perturbatively expand in the amplitude of the non-symmetric components of the magnetic field and calculate the drift surfaces and the adiabatic invariant $\Jpl$ for this model magnetic field. Finally, we discuss the relationship between the drift surfaces and surfaces of constant $\psi$ and $q$.

\subsection{An exact vacuum magnetic field }
Let $(X, Y, Z)$ denote a point in a flat toroidal shell (a topological torus). The coordinates $Y$ and $Z$ denote the toroidal and the poloidal angles respectively, and the coordinate $X$ labels the shells. The metric of this space is Euclidean. The main advantage of working in a topological torus is that it simplifies the analysis due to the absence of toroidal curvature effects but retains fully the complexities associated with the double periodicity of the torus. 

The magnetic field is given by
\begin{eqnarray}
{\bf B}= \dl \Phi, \quad \text{where} \quad
\Phi= Y + \delta \sin Y \cos Z \cosh \sqrt{2}X,
\label{phi_def}
\end{eqnarray}
and $\delta$ is a constant.
The magnetic field can also be defined in terms of Clebsch coordinates $(\psi, \alpha)$ as 
\begin{eqnarray}
{\bf B}= \dl \psi \times \dl \alpha,
\end{eqnarray}
with the following $(\psi,\alpha)$ characteristics 
\begin{subequations}
\begin{align}
    \psi(X,Y,Z)&=\sqrt{2\sinh{(\sqrt{2}|X|})}\sin{|Z|}
    \label{psi_exp}\\
    \alpha(X,Y,Z)&=\delta \sqrt{\sinh{(\sqrt{2}|X|})}\cos{Y}\label{alpha_exp}+\sigma_z\int^{|X|}_{|X_0|}\dfrac{dX'}{\sqrt{2\sinh{(\sqrt{2}X')}-\psi^2}} \\
    \text{where,}\quad  \sigma_Z &= \text{sign}(\cos{Z}),\: X_0 =
 \dfrac{1}{\sqrt{2}}\sinh^{-1}{\lbr\frac{\psi^2}{2}\rbr}.\nonumber
\end{align}
\label{psi_alpha}
\end{subequations}
The function $\psi$ defined above is a single-valued function, whereas the function $\alpha$ is not because of the second term in \eqref{alpha_exp}. Note that, instead of $(\psi,\alpha)$ we can also choose any functions of $(\psi,\alpha)$, for example, $(\hat{\psi}=\psi^2/2, \hat{\alpha}=\alpha/\psi)$ as the Clebsch coordinates.

\begin{figure}
    \centering
    \includegraphics[width=0.6\textwidth]{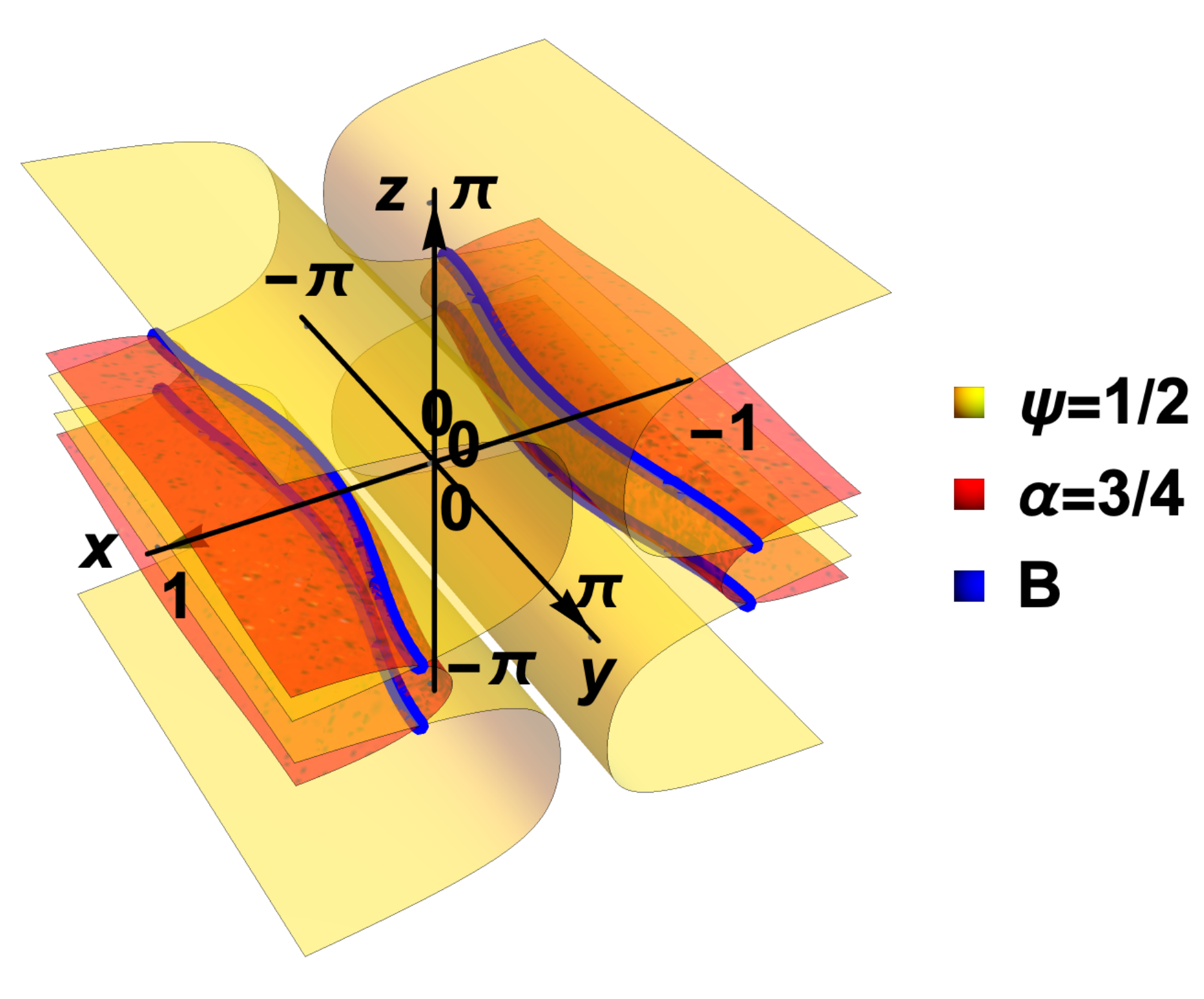}
    \caption{Clebsch coordinates $\psi$(yellow),$\alpha$(orange), and the magnetic field lines (blue)} satisfying \eqref{phi_def}-\eqref{psi_alpha}. Note the even parity (reflection symmetry) in all the three coordinates.  
    \label{psi3D}
\end{figure}

The components of the magnetic field are 
\begin{subequations}
\begin{align}
B_X&= \sqrt{2}\delta \sin Y \cos Z \sinh \sqrt{2}X.\\
B_Y&= 1+\delta \cos Y \cos Z \cosh \sqrt{2}X. \\
B_Z&= -\delta \sin Y \sin Z \cosh \sqrt{2}X. 
\end{align}
\label{BxByBz}
\end{subequations}
We shall use $\delta$ as a convenient expansion parameter in this work. To lowest order the magnetic field is
\begin{eqnarray}
B_X^{(0)}=0,\:\:B_Y^{(0)}=1,\:\:B_Z^{(0)}=0.
\label{B_0}
\end{eqnarray}
The above equations show that the lowest order magnetic field is of uniform strength $(B^{(0)}=|{\bm B}^{(0)}|=1)$ and is purely in the $Y$ (toroidal) direction.     

To describe the motion of the charged particles, we shall need to parametrize the magnetic field lines. Since the particles follow the magnetic field lines to the lowest order in the gyroradius, the coordinates used to describe the field lines will also determine the lowest order particle trajectories. The equations describing the magnetic field lines are given by
\begin{eqnarray}
\frac{dX}{B_X}=\frac{dY}{B_Y}=\frac{dZ}{B_Z}.
\end{eqnarray}
Therefore,
\begin{subequations}
\begin{eqnarray}
  \frac{dX}{dY} = \frac{B_X}{B_Y}=\frac{\sqrt{2}\delta \sin Y \cos Z \sinh \sqrt{2} X}{1+\delta \cos Y \cos Z \cosh \sqrt{2} X}\\
 \frac{dZ}{dY} = \frac{B_Z}{B_Y}=\frac{-\delta \sin Y \sin Z \cosh \sqrt{2} X}{1+\delta \cos Y \cos Z \cosh \sqrt{2} X}.
 \end{eqnarray}
 \label{fieldline_eqns}
\end{subequations}
To simplify algebra we introduce the following variables 
\begin{align}
    x=\sqrt{2} X,\: y=Y,\:z=Z,\: \eta =\cos{Y},
    \label{smallx_def}
\end{align}
and rewrite \eqref{fieldline_eqns} as
\begin{subequations}
\begin{eqnarray}
\frac{dx}{d\eta}&=& \frac{-2\delta \cos z \sinh x}{1+\delta \eta \cos z \cosh x}  \\
\frac{dz}{d\eta}&=& \frac{\delta \sin z \cosh x }{1+\delta \eta \cos z \cosh x}.
\end{eqnarray}
\label{fieldline_eta}
\end{subequations}
Equations \eqref{fieldline_eta} are difficult to solve exactly, so series expansions in $\delta$ will now be constructed. 
From the structure of \eqref{fieldline_eta} we find that  $x$ and $z$ can be expanded in powers of $\delta$ as follows
\begin{eqnarray}
x=x_0+ \delta\: x_1\eta+ \delta ^2 x_2\eta^2+ \cdots = \sum_n x_n\delta^n \eta^n\\
z=z_0+ \delta\: z_1\eta+ \delta ^2 z_2\eta^2+ \cdots = \sum_n z_n\delta^n \eta^n, \nonumber
\end{eqnarray}
and substituting in \eqref{fieldline_eta} we obtain

\begin{eqnarray}
\frac{dx_0}{d\eta}+ \delta\frac{dx_1}{d\eta}+ O(\delta^2)= -2\delta \sinh x_0+ O(\delta^2) \nonumber\\
\frac{dz_0}{d\eta}+ \delta\frac{dz_1}{d\eta}+ O(\delta^2)= \delta \cosh x \sin z_0+ O(\delta^2)
\label{xzeqn_exp}
\end{eqnarray}

Using \eqref{xzeqn_exp} we determine $x^{(n)}$ and $z^{(n)}$ order by order. In particular, 
 \begin{eqnarray}
  \frac{d x_0}{d\eta}=0, \quad
\frac{d z_0}{d\eta}=0,
\end{eqnarray}
which shows that both $x_0$ and $z_0$ must be constants. This is expected since the lowest order magnetic field \eqref{B_0} is purely in the $Y$ (toroidal) direction.
The rest of $(x^{(n)},z^{(n)})$ are also obtained as functions of $(x_0,z_0)$.


The quantity $q=\oint dl/B$ for this magnetic field  \citep{weitzner_sengupta2019exact}, obtained as a function of $(x_0,z_0)$ is
\begin{align}
    \frac{q}{2\pi}= 1+\frac{1}{2}\cos^2{z_0}\cosh^2{x_0}\:\delta^2 +O(\delta^4)\, .
    \label{qdef}
\end{align}

\subsection{Evaluation of the omnigeneity constraint}
\label{omni_sec}
The equation for $J_{\parallel}$ (equation (5)) can be rewritten as 
\begin{eqnarray}
\Jpl=\oint v_{\parallel}d\ell = \oint \vpl \frac{dy}{B_y}.
\end{eqnarray}
Since, lowest order $B$ is 1, the above expression is approximately equivalent to
\begin{eqnarray}
 \Jpl\approx \oint \vpl dy =
\oint \sigma\sqrt{2(\mathcal{E}-\mu B)} dy.
\end{eqnarray}
To evaluate $\Jpl$, we expand $B$ as follows
\begin{align}
B =& \sqrt{B_X^2+B_Y^2+ B_Z^2}= 1 + \delta \cos y \cos z_0\cosh x_0+ O(\delta^2).
\end{align}
Therefore,
\begin{align}
&B = 1+ \delta B_1 + O(\delta^2) \nonumber\\
\text{where}, \quad &\delta B_1=\delta_0\cos y \quad \text{and}\quad \delta_0= \delta \cosh{x_0}\cos z_0.
\label{B1_def}
\end{align}
Since $B$ is sinusoidal in $y$ to this order, $\Jpl$ is identical to the action coordinates of a simple pendulum \citep{brizard2011action_pendulum}. We present the details of the calculation in appendix \ref{appnedix_Jpl}.

We find that the lowest order $\Jpl$ for both trapped and circulating particles are functions of only $|\delta_0|$. From \eqref{qdef} we find that,
\begin{align}
   \delta_0=\delta_0(q) \approx \sqrt{2}\sqrt{\frac{q}{2\pi}-1}.
    \label{delta0q}
\end{align}
Hence, $\Jpl=\Jpl(q)$ to lowest order. Since $q=q(\psi,\alpha)$, $\Jpl=\Jpl(\psi,\alpha)$ for both trapped and circulating particles. This shows that the closed field line omnigeneity condition \eqref{omni_closed} is satisfied but not the omnigeneity condition \eqref{omnigeneity} for field lines with flux-surfaces.

\subsection{Evaluation of the QS constraint}
\label{QS_sec}
We have seen that the closed field line omnigeneity condition is satisfied to lowest order. We now present the calculation of the QS constraints \eqref{QS2_q} to the lowest nontrivial order. We begin with the QS condition \eqref{QS1_psi}. We note from \eqref{BxByBz},\eqref{smallx_def} and \eqref{B1_def} that 
\begin{align}
   &\BD = \frac{\del}{\del y} + \delta \lbr \cos y \cos z \cosh{x}\frac{\del}{\del y}+
    2\sin y \lbr\cos z 
    \sinh{x}\frac{\del}{\del x}-\sin z \cosh{x}\frac{\del}{\del z}\rbr \rbr \nonumber\\
  &\BD B = \delta \frac{\del B_1}{\del y} +O(\delta^2),
\quad \quad\B \times \dl B =\delta \dl y \times \dl B_1+ O(\delta^2).
\label{BD_BDB_BxDB}
\end{align}
From \eqref{psi_exp} and \eqref{BD_BDB_BxDB} it follows that
\begin{align}
 \cF=   \frac{\B\cdot \dl \psi \times \dl B}{\BD B}=\cot y\frac{\cosh x  \cos {z} \left(2 \tanh ^2 x + \tan^2 z\right)}{\sqrt{\sinh  x}}+O(\delta)
    \label{QSconstraint_not}
\end{align}
The factor of $\cot{y}$ implies that $\BD \cF \neq 0$. Hence the QS condition \eqref{QS1_psi} is not satisfied. 

However, the QS condition for closed field lines \eqref{QS2_q} is satisfied since
\begin{align}
    \frac{\cG}{\delta^2}= \frac{\B\cdot \dl (q/\delta^2) \times \dl B}{\BD B}= \frac{\dl y \times \dl (q/\delta^2) \cdot \dl (\delta_0(q) \cos{y})}{\del_y (\delta_0(q) \cos{y})}+O(\delta)= O(\delta)
    \label{QSconstraint_q_yes}
\end{align}
The division of $\cG$ by $\delta^2$ was done because $\dl q$ is of $O(\delta^2)$. The lowest order term in \eqref{QSconstraint_q_yes} vanishes, showing that the closed field line QS condition is indeed satisfied to lowest order. Comparing \eqref{QSconstraint_not} and \eqref{QSconstraint_q_yes}, we see that the former is not QS while the latter is only because $\del_\alpha q(\psi,\alpha)\neq 0$. 

QS can also be inferred from the expression for the field strength,
\begin{align}
    B=1+\cos{y}\:\delta_0(q) + O(\delta^2).
\end{align}
To first order, $B$ depends only on one angle, $y$, when $q$ is used as a coordinate. Hence, $B$ is QS. We present a more detailed description of the QS calculation in appendix \ref{appendix_QS_higherorder}.

\subsection{Comparison of drift, $\psi$ and $q$ surfaces}
\label{for_comparisons_sec}
The drift surface of a charged particle obtained after averaging out the gyromotion and the bounce motion is a surface on which $\Jpl$ is constant \citep{gardner1959adiabatic}. We shall now compare the drift surfaces with the constant $\psi$ and $q$ surfaces. 

\begin{figure}
\subfigure[ $\psi$ as a function of $(x,z)$]{
\includegraphics[width=0.45\linewidth]{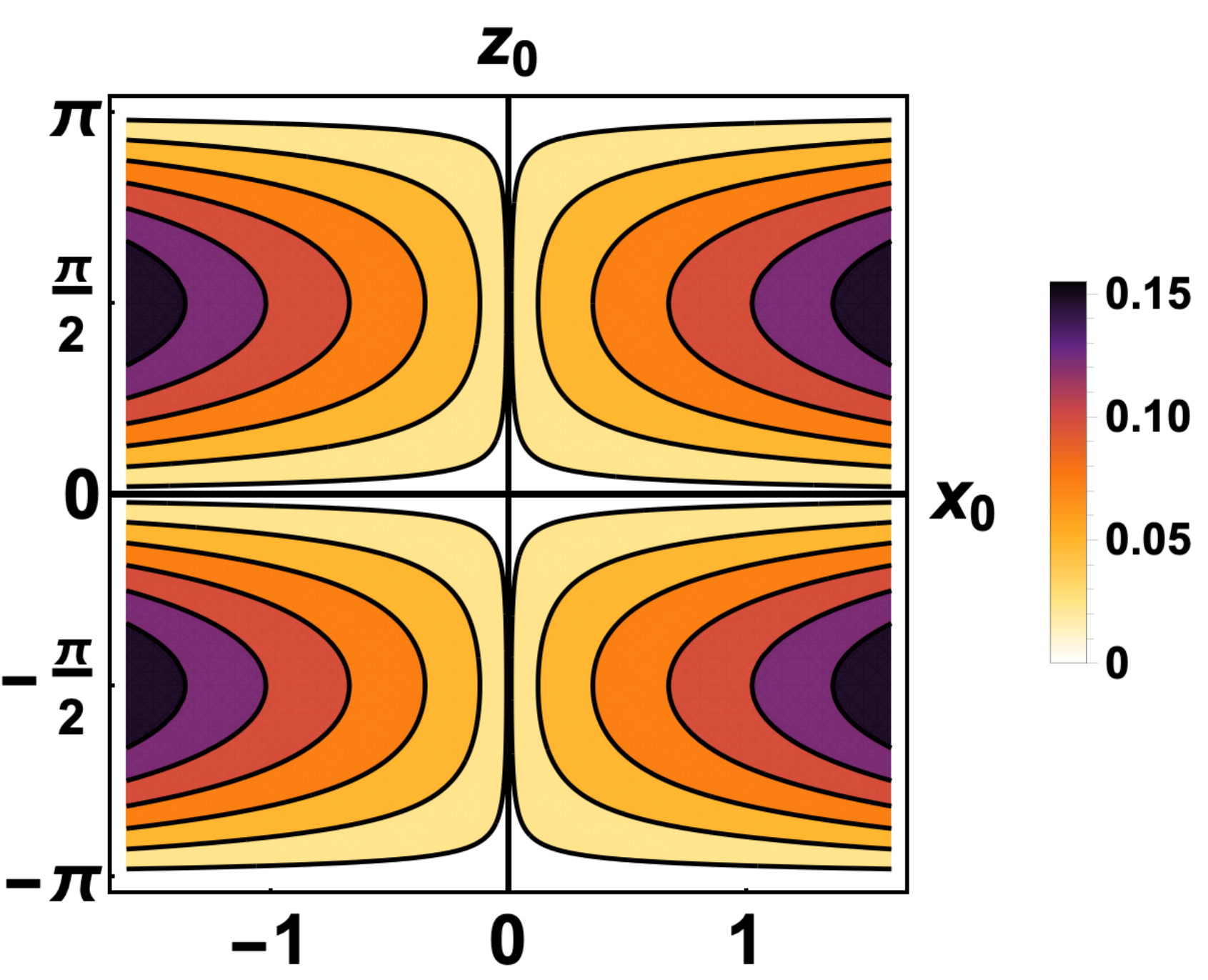}\label{fig:psi_contours}
}\hfill
\subfigure[$q/(2\pi)-1$ as a function of $(x_0,z_0)$]{\includegraphics[width=0.45\linewidth]{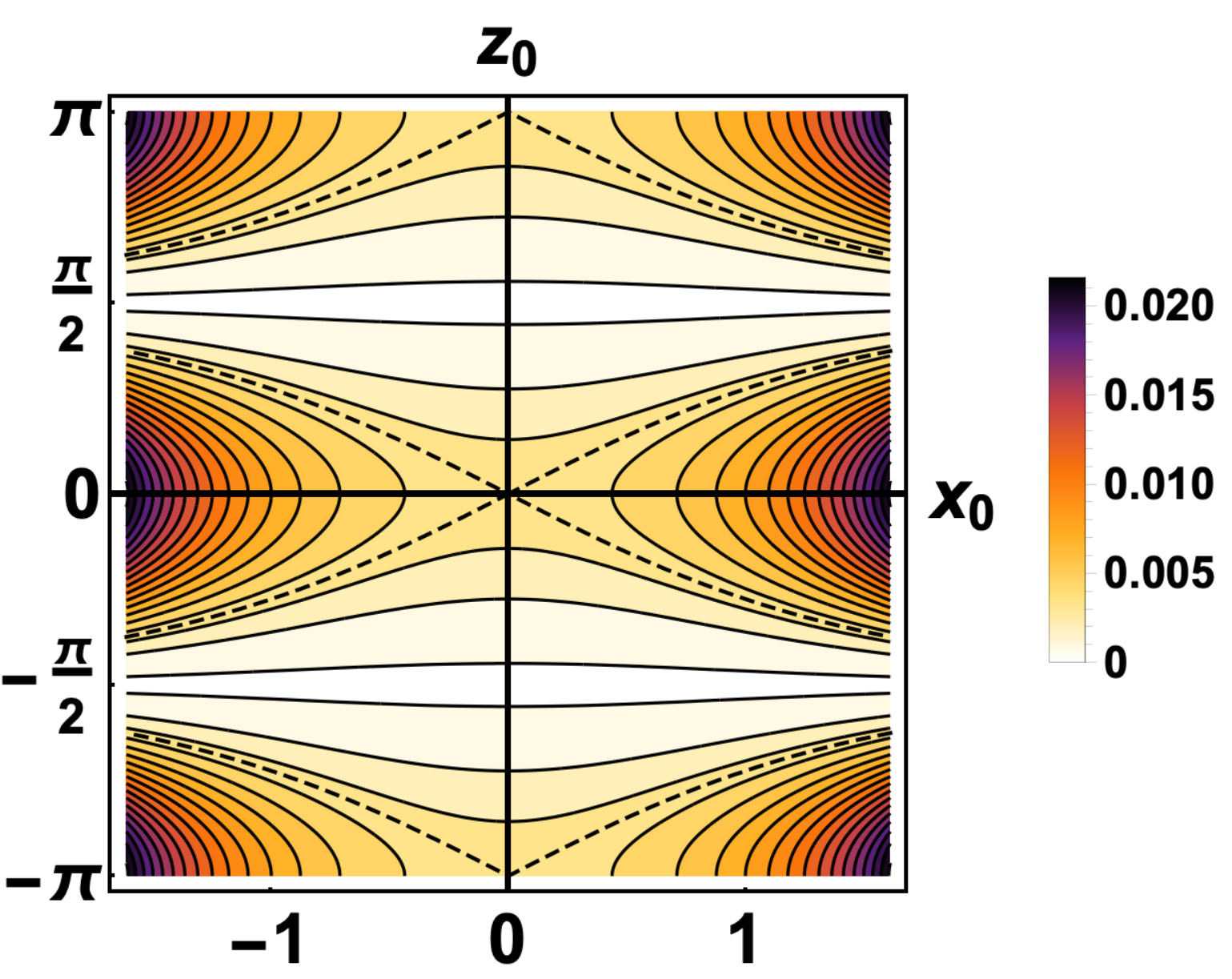}\label{fig:q_contours}}
\caption{\label{fig:psi_q_contours} Contours of $\psi$ and $q$ for $\delta=0.08$.  The dotted lines in \ref{fig:q_contours} is the separatrix $\cos{z_0}\cosh{x_0}=\pm 1$.}
\end{figure}

Figure \ref{fig:psi_q_contours} shows the constant $\psi$ and $q$ contours. The contours of $q$ and $\psi$ do not coincide since $q=q(\psi,\alpha)$.  From the contours of the constant $\psi$ surfaces in figure \ref{fig:psi_contours} and equation \eqref{BxByBz}, we see that the magnetic field has X-points at $(x=0,z=0)$ and $(x=0,z=\pm\pi)$, where $B_X=B_Z=0$ but $B_Y\neq 0$. In a topological torus, $z=-\pi$ and $z=\pi$ are identified. Hence, there are only two X-points. The difference between $(x_0,z_0)$ and $(x,z)$ can be ignored for the lowest order description that we are interested in, but must be taken into account for higher orders.

\begin{figure}
\subfigure[ $\Jpl^{TP}$ for TP near $z=0$ and $q$ surfaces ]{
\includegraphics[width=0.45\linewidth]{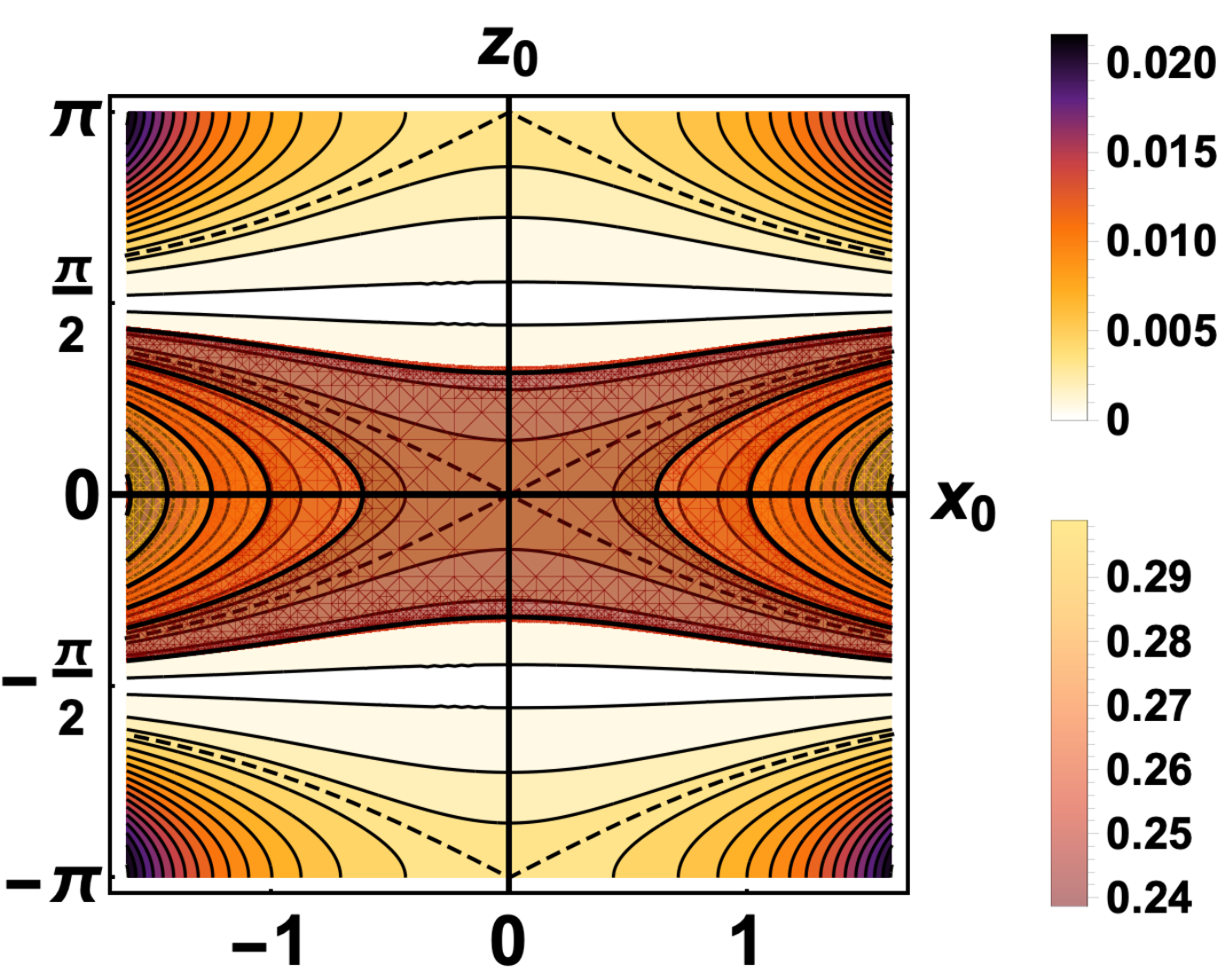}\label{fig:q_JPl_z0}
}\hfill
\subfigure[ $\Jpl^{TP}$ for TP near $z=\pm \pi$ and $q$ surfaces]{\includegraphics[width=0.45\linewidth]{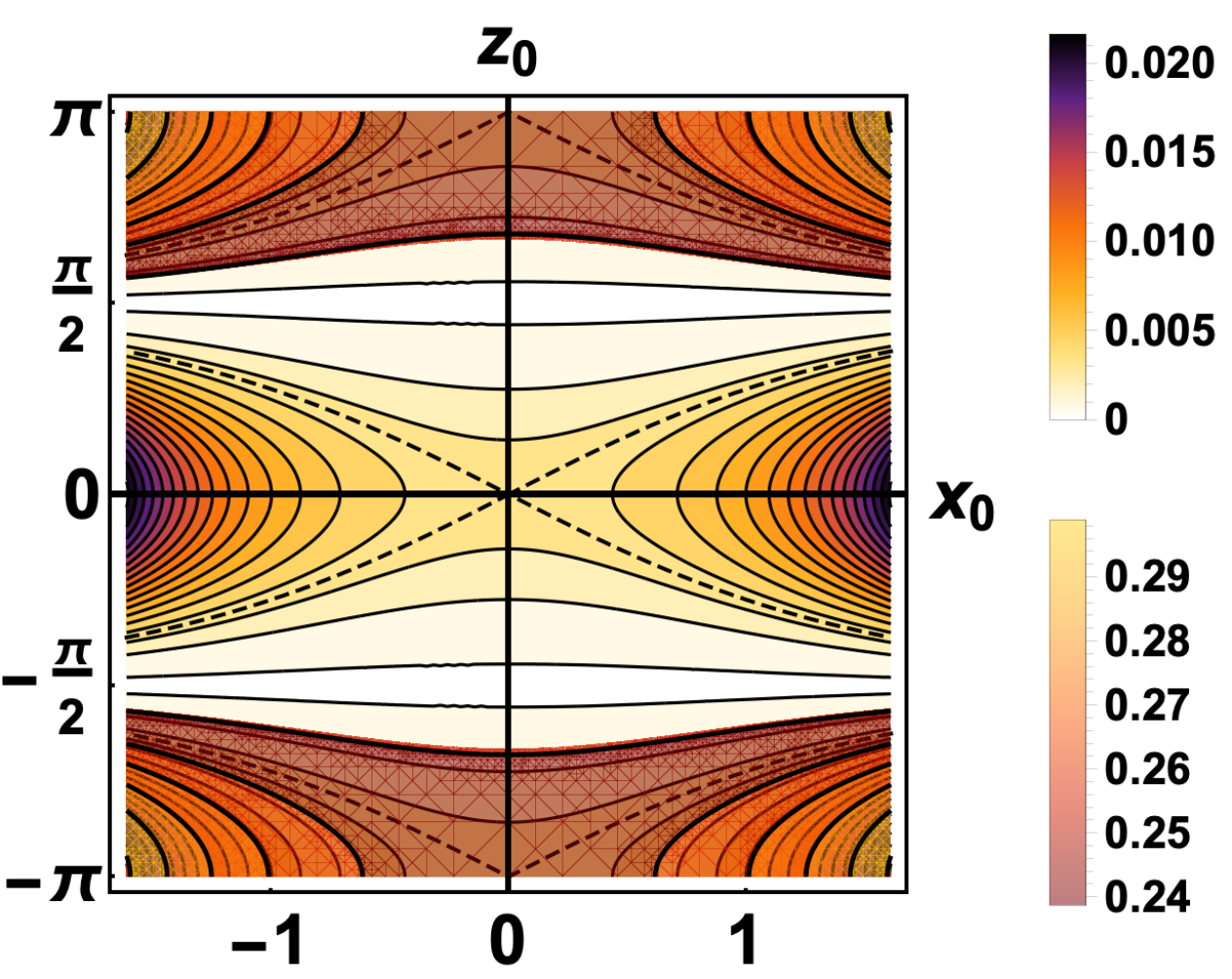}\label{fig:q_JPl_zpi}}
\caption{\label{fig:q_JPl_contours} Drift surfaces (brown shades) of trapped particles superposed on $q$ surfaces for $\ep=0.08$.}
\end{figure}

\begin{figure}
\subfigure[ $\Jpl$ for TP near $z=0$ and $\psi$ surfaces ]{
\includegraphics[width=0.45\linewidth]{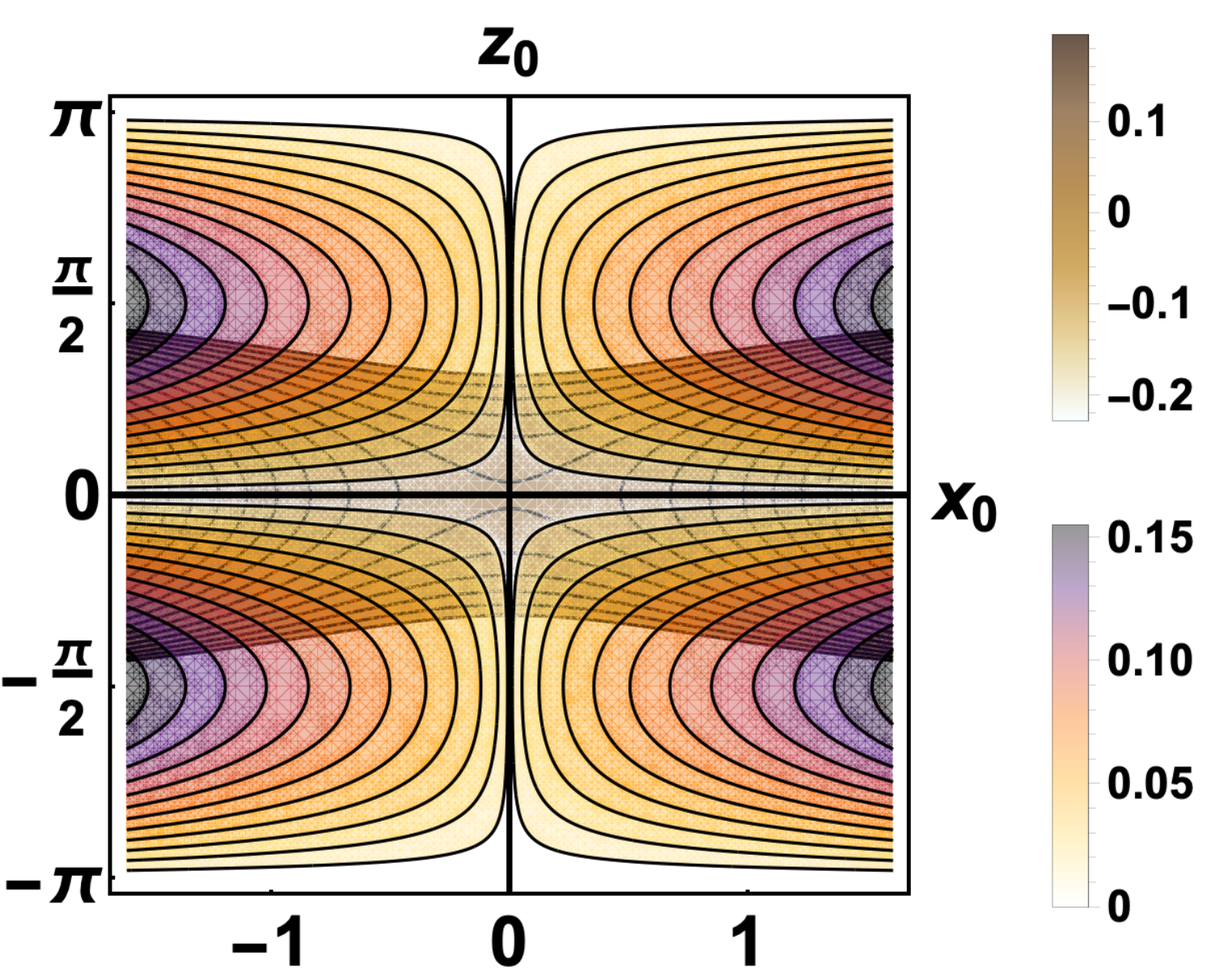}\label{fig:psi_JPl_z0}
}\hfill
\subfigure[ $\Jpl$ for TP near $z=\pm \pi$ and $\psi$ surfaces]{\includegraphics[width=0.45\linewidth]{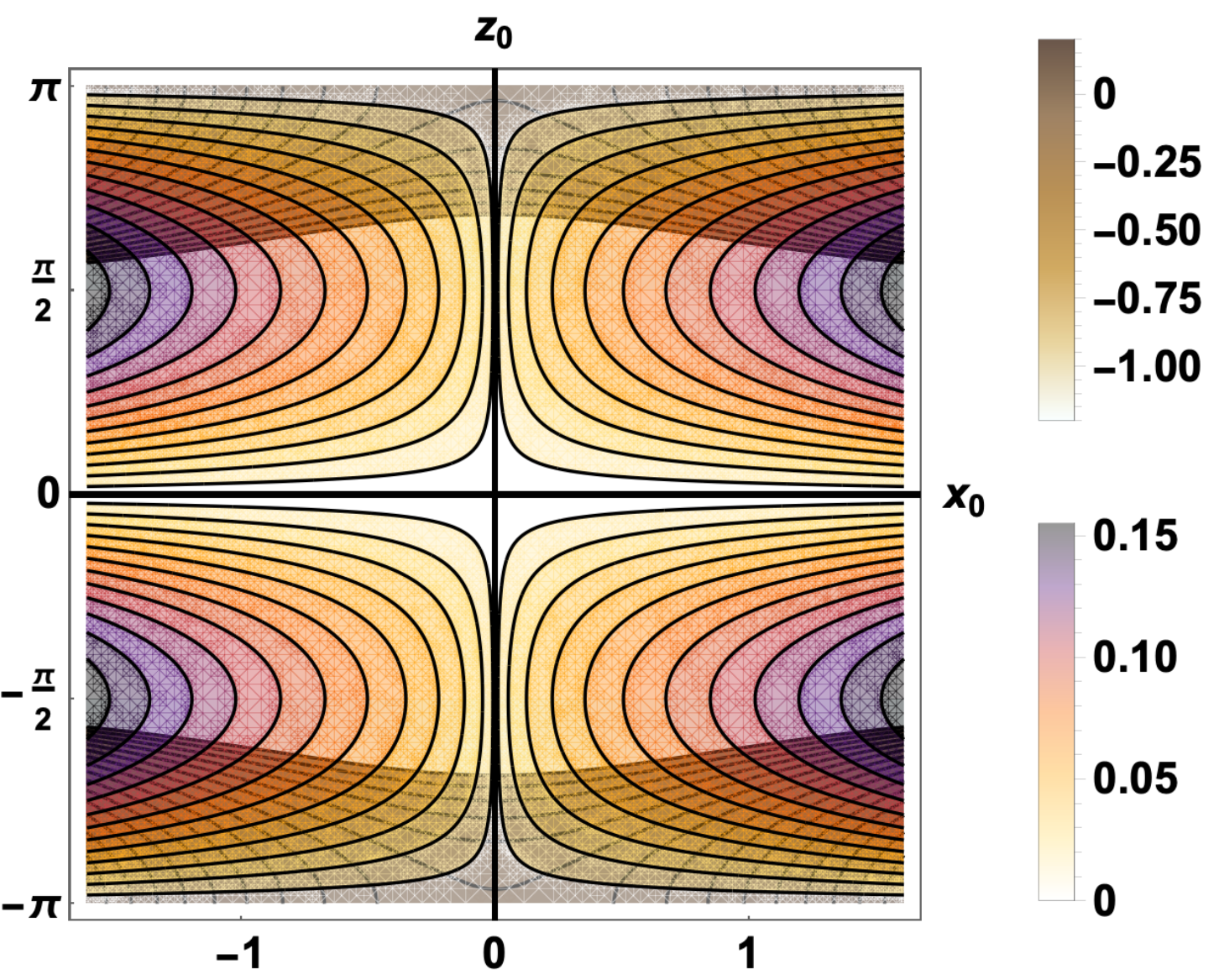}\label{fig:psi_JPl_zpi}}
\caption{\label{fig:psi_JPl_contours} Drift surfaces (brown shades) of trapped particles superposed on $\psi$ surfaces for $\ep=0.08$}
\end{figure}

We have shown in section \ref{omni_sec} and appendix \ref{appnedix_Jpl} that $\Jpl$ and $q$ surfaces coincide to lowest order. In figure \ref{fig:q_JPl_contours} we superpose the contours of constant $\Jpl$ for trapped particles on top of contours of constant $q$. Figure \ref{fig:q_JPl_z0} and \ref{fig:q_JPl_zpi} show the drift surfaces of particles trapped near $z_0=0$ and $z=\pm \pi$. In figure \ref{fig:psi_JPl_contours} we compare the trapped particle $\Jpl$ with $\psi$ surfaces. It is clear that the particles will drift on a constant $q$ surfaces and not constant $\psi$ surfaces. If $q$ surfaces are closed, then all the particles will be confined. To the order of accuracy considered here, $q$ surfaces are not closed. However, obtaining closed $q$ surfaces is possible, as shown in \cite{weitzner_sengupta2019exact}. 

We, therefore, conclude that the model magnetic field \eqref{BxByBz}, which was not optimized to satisfy any constraint, turns out to be QS and hence omnigeneous to lowest order in $\delta$, provided the closed field line criteria \eqref{omni_closed} and \eqref{QS2_q} are used. The drift surfaces coincide with the pressure and $q$ surfaces and not the $\psi$ surfaces. Neither omnigeneity nor QS holds if the regular expressions \eqref{omnigeneity} and \eqref{QS2_psi}, which are valid for irrational rotational transform, are used. 

\section{Discussion}\label{sec:discussion}
Modern stellarators experiments like the high-$\iotabar$ configuration of W7-X \citep{andreevaW7Xvacuum,klinger2019overview} can have very low global magnetic shear and a rotational transform close to a low-order rational number. One can regard such a configuration as a composition of closed-field lines corresponding to the low-order rational iota, and a small part which generates the flux surfaces \citep{taylor_hastie,cary_Hedrick_1988orbits}. The adiabatic invariant in such a case turns out to be the canonical angular momentum averaged along the closed field line \citep{taylor_hastie,cary_Hedrick_1988orbits}. In principle, the closer the rotational transform is to a rational number, the better the closed field line approximation should be. However, there are crucial differences between MHD equilibrium and MHD stability for rational vs. irrational rotational transform \citep{grad1971plasma,grad1973magnetofluid,spies1974nonlinear_stability_closed}. In particular, for MHD stability, the limit of zero magnetic shear might be singular, and the exact closed field line system can be very different from a neighboring system with very small shear \citep{spies1979low_report,nelson_spies_1974_suff_MHD_closed}.

On the other hand, various asymptotic expansions \citep{weitzner2016expansions,sengupta_erin2018,sengupta_weitzner2019_lowshear_vacuum} provide hints that vacuum and ideal MHD equilibria with flux surfaces and continuous rotational transform and pressure profiles might exist provided that the rotational transform is close to a low-order rational number, the magnetic field shear is low, and the boundary is chosen self-consistently. The jump from a closed-field line MHD equilibrium with exactly zero shear to a neighboring equilibrium with minimal magnetic shear might not be discontinuous \cite{weitzner2016expansions}. It is, therefore, useful to investigate the omnigeneity and quasisymmetry properties of closed field line systems.

In this work, we have investigated the radial localization of particles in a magnetic field that has zero magnetic shear and all closed lines. We formulated the radial localization of particles and the omnigeneity and quasisymmetry constraints in terms of the pressure surfaces and not the flux-surfaces. Our definitions are physically motivated since any confined equilibria should have closed pressure surfaces, and a configuration where particles stay localized near such closed surfaces should be ideal. However, in closed field line systems, the pressure is a function of $q=\oint dl/B$, which is, in general, dependent on both the Clebsch variables. Hence, the standard formulation of omnigeneity and quasisymmetry based on irrational rotational transform must be modified to take closed field lines into account. A closed field line configuration where pressure surfaces (constant $q$) and drift surfaces (constant $\Jpl$) coincide, if it exists, can be shown to be interchange-stable \citep{benDaniel1965nonexistence,taylor_hastie}. Differences between rational and irrational rotational transform also show up in the adiabatic invariants for circulating and trapped particles \citep{grad1967toroidal,burby_Kallinikos_MACKAY2019_Math_QS}. For ergodic field lines with flux-surfaces, the drift orbits of circulating particles are always radially localized, unlike closed field line systems. 

We have then discussed drift orbits in a model magnetic field with closed field lines obtained recently \citep{weitzner_sengupta2019exact}. Expanding in a small parameter that measures the lack of symmetry of the magnetic fields, we show that the drift surfaces to lowest order are functions of $q$. The model configuration is both omnigeneous and quasisymmetric according to the proposed constraints for closed field line systems, but not any of these according to the general definitions valid for irrational rotational transform. The analysis presented here suggests that optimization codes can easily miss out configurations where the drift surfaces coincide with the pressure surfaces in closed field line configurations simply because the definitions of the omnigeneity and QS constraints are too restrictive.

The drift surfaces characterize the collisionless cross-field motion of particles. Our results show that in closed field line systems, the cross-field drift from both circulating particles and trapped particles can be significant. Also, in closed field line systems, the function $q$ can play a vital role in determining the drifts of the particles. In numerically optimizing stellarators for QS and fast-particle transport with low-shear and near rational rotational transform configurations, the function $q$ could be relevant.

In the future, we shall consider extending the closed field omnigeneity and QS conditions to include the effects of small shear following \cite{cary_Hedrick_1988orbits,taylor_hastie}. The class of exact vacuum solutions obtained in \cite{weitzner_sengupta2019exact} is very large, and we can investigate which configuration, if any, can exactly or approximately satisfy omnigeneity or QS in a volume.

\noindent{\bf Acknowledgements}
This work was done as part of the NYU Courant Institute Girls' Science, Technology, Engineering, and Mathematics  (NYU GSTEM) Summer Program. EE would like to thank the GSTEM program for the opportunity; specifically, Catherine Tissot, for the work she put into it. In addition, we would like to thank Hannah Boland for being a supportive tutor. EE would also like to thank everyone in the math department at Courant 
for the insightful conversations that were shared, especially Jeff McFadden, for all the books. 
This research was partly
funded by the US DOE grant no. DEFG02-86ER53223.
\appendix

\section{Quasisymmetry and periodicity of $B$}
\label{appendix_QS_periods}
Quasisymmetry can be formulated as a periodicity condition on $B$ \citep{helander2014theory}. For ergodic field lines with flux surfaces, the QS requirement is that the period is constant on the surface. Hence,
\begin{align}
    B(\psi,\alpha,l+L(\psi))=B(\psi,\alpha,l).
\end{align}
We now prove that the same property is true in a closed field line system with the replacement of $\psi$ by $q$. We start with the ideal MHD equilibrium conditions \eqref{weitzner_MHD_formalism} with $(q,\alphabar)$ as the Clebsch potentials, and the QS condition \eqref{QS2_q}
\begin{subequations}
\begin{align}
\B=\dl q \times \dl \alphabar = \frac{\dl q \times (\dl\Phi \times \dl q)}{|\dl q|^2}, \quad \dl q\times \dl \alphabar \cdot \dl \lbr \frac{\dl q\cdot \dl \Phi}{|\dl q|^2}\rbr&=-p'(q).
\\
  \lbr \B \times \dl q -\cG(q)\B\rbr \cdot \dl B &=0
  \label{QSPDE_q}
\end{align}
\end{subequations}
Using $$\B \times \dl q = \dl \Phi \times \dl q, \quad \B = \dl q\times \dl \alphabar,$$
and $(q,\alphabar,\Phi)$ coordinate system, we rewrite \eqref{QSPDE_q} as
\begin{align}
\lbr \frac{\del}{\del \alpha} -\cG(q) \frac{\del}{\del \Phi}\rbr B =0.
    \label{QSPDE_characteristics}
\end{align}
It follows from \eqref{QSPDE_characteristics} that $B$ must be of the form
\begin{align}
B(q,\alphabar,\Phi)=\cB(\Phi + \cG(q)\alphabar).
    \label{Bform}
\end{align}
For fixed $q$ and $\alphabar$, the periodicity of $B$ is determined by the periods of $\Phi$, which from \eqref{PhiZetabox} are functions of $q$, and therefore fixed on a constant $q$ surface. Hence, the QS periodicity requirement is satisfied. The exact same analysis is valid for ergodic field lines with $(q,\alphabar)$ replaced by $(\psi,\alpha)$.

\section{Details of the calculation of $\Jpl$}
\label{appnedix_Jpl}
We present here the details of the calculation of the drift surfaces for trapped and circulating particles. Trapped particles (TP) occupy the region where the magnetic field strength has a minimum. From \eqref{B1_def}, we see that the minima occur when
\begin{align}
    \del^2_y B=-\delta_0(x_0,z_0) \cos{y} +O(\delta^2)>0.
    \label{Bmin}
\end{align}
Therefore, particles are trapped near $y=\pm \pi$ for $\delta_0(x_0,z_0)>0$, and $y=0$ for $\delta_0(x_0,z_0)<0$.  With the choice $\delta>0$, it follows from \eqref{B1_def} that
\begin{align}
    \delta_0(x_0,z_0) = 
    \begin{cases}
    \delta \cos{z_0}\cosh{x_0}<0, \quad
    -\pi\leq z_0\leq -\pi/2\:\: \text{and} \:\: +\pi/2\leq z_0\leq \pi\\
     \delta \cos{z_0}\cosh{x_0}>0, \quad -\pi/2\leq z_0\leq +\pi/2
    \end{cases}
\end{align}
We begin with the case where the particles are trapped near $y=0$, i.e. $\delta_0<0$. 
To lowest nontrivial order, $J_{\parallel}$ is given by
\begin{eqnarray}
J_{\parallel}\approx \oint \sigma \sqrt{2(\mathcal{E}-\mu(1- \delta_r\cos y))}dy,
\end{eqnarray}
where 
\begin{align}
    \delta_r=-\delta_0 = -\delta \cos{z}\cosh{x} \geq 0.
\end{align}

For trapped particles (TP) the $\Jpl$ integral is carried out between the bounce points $-y_b$ and $+y_b$. Taking into account that $\sigma$ can be $\pm 1$ for trapped particles,
\begin{eqnarray}
J^{TP}_{\parallel}= 2 \int_{-y_b}^{+y_b}\sqrt{2(\mathcal{E}-\mu(1- \delta_r\cos y))}\:dy.
\end{eqnarray}
At the endpoints, 
\begin{eqnarray}
\cE/\mu =B(y_b)= (1-\delta_r\cos y_b).
\end{eqnarray}
Solving for the bounce point we get,
\begin{align}
\sin^2 \frac{y_b}{2}=  \frac{\mathcal{E}/\mu -(1-\delta_r)}{2\delta_r},
\end{align}
which implies that the trapped particles localized near $y=0$, have $\cE/\mu$ ratio bounded between
\begin{eqnarray}
1-\delta_r \le \frac{\mathcal{E}}{\mu}\le 1+\delta_r.
\end{eqnarray}
The $J_{\parallel}$ integral takes the following form
\begin{eqnarray}
J^{TP}_{\parallel}= 2 \sqrt{\mu \delta_r}
\int_{-y_b}^{y_b} \sqrt{\sin^2 \frac{y_b}{2}-\sin^2 \frac{y}{2}}dy \:.
\end{eqnarray}
For deeply trapped particles $\sin y\approx y$,  therefore,
\begin{eqnarray}
J^{TP}_{\parallel} = 2\pi \lbr\frac{y_b}{2}\rbr^2 \sqrt{\mu \delta_r}\;.
\end{eqnarray}
In general $\Jpl$ can be written in terms of elliptic integrals \citep{brizard2011action_pendulum},
\begin{eqnarray}
J^{TP}_{\parallel} = 8 \sqrt{\mu \delta_r} \lbr E(k^2)-(1-k^2)K(k^2) \rbr,\quad k=\sin{\frac{y_b}{2}}.
\label{JplTP_gen}
\end{eqnarray}
For particle trapped near $y=\pm \pi$, we shift $y$ to $y'=y\pm \pi$ such that $\cos{y}=-\cos{y'}$. Also,
\begin{align}
    \sin^2{\frac{y'_b}{2}}= \frac{\cE/\mu -(1-\delta_0)}{2\delta_0}, \quad 1-\delta_0\leq \frac{\cE}{\mu} \leq 1+\delta_0.
\end{align}
Both of these cases can be represented by \eqref{JplTP_gen} with $\delta_r$ replaced by $|\delta_0|$.

Similarly, for circulating particles (CP), carrying out the integration from $0$ to $2\pi$, we obtain
\begin{align}
    J^{CP}_{\parallel} = 4\sigma \sqrt{\mu |\delta_0|} \: k E(1/k^2), \quad \text{where}\quad k^2=\frac{\cE/\mu-(1-|\delta_0|)}{2|\delta_0|}.
\end{align}

\section{Details of the QS calculation}
\label{appendix_QS_higherorder}
The function $q$ is a function of $(x_0,z_0)$, it is convenient to rewrite the QS condition \eqref{QS2_q} in terms of $(x_0,z_0)$. Along the field line characteristics \eqref{fieldline_eta}
\begin{align}
    x=\sum_{n=0}^\infty (\delta \eta)^n x^{(n)}(x_0,z_0),\quad y =\cos^{-1}{\eta}\, ,\quad z=\sum_{n=0}^\infty (\delta \eta)^n z^{(n)}(x_0,z_0)
\end{align}
Taking the gradient with respect to the $(x,y,z)$ coordinate system we get
\begin{align}
\begin{pmatrix}
\dl x\\
\\
\dl y\\
\\
\dl z
\end{pmatrix} =
 \begin{pmatrix}
\dfrac{\del x}{\del x_0} & \dfrac{\del x}{\del \eta} &\dfrac{\del x}{\del z_0} \\
\\
0 & \dfrac{dy}{d\eta}& 0\\
\\
\dfrac{\del z}{\del x_0} & \dfrac{\del z}{\del \eta} & \dfrac{\del z}{\del z_0} 
\end{pmatrix}
\begin{pmatrix}
\dl x_0\\
\\
 \dl \eta \\
 \\
\dl z_0
\end{pmatrix}
\end{align}
which can be inverted to obtain 
\begin{align}
\begin{pmatrix}
\dl x_0\\
 \\
 \dl \eta \\
 \\
\dl z_0
\end{pmatrix} = \dfrac{1}{\{x,z\}_{(x_0,z_0)}}
 \begin{pmatrix}
\dfrac{\del z}{\del z_0} & \dfrac{d \eta}{d y}\{x,z\}_{(z_0,\eta)} & -\dfrac{\del x}{\del z_0} \\
0 & \dfrac{d \eta}{d y} & 0
\\
-\dfrac{\del z}{\del x_0} & \dfrac{d \eta}{d y} \{x,z\}_{(\eta,x_0)} &\dfrac{\del x}{\del x_0} 
\end{pmatrix}
\begin{pmatrix}
\dl x\\
\\
 \dl y\\
 \\
\dl z
\end{pmatrix}
\end{align}
where $\{x,z\}_{(a,b)}$ denotes the Poisson bracket of $(x,z)$ with respect to $(a,b)$. it follows
\begin{align}
    \dl q(x_0,z_0) =& \dfrac{1}{\{x,z\}_{(x_0,z_0)}}\left( \{q,z\}_{(x_0,z_0)}\dl x+ \{x,q\}_{(x_0,z_0)} \dl z\right.\nonumber\\
    &+\dfrac{d\eta}{dy}\left. \lbr \dfrac{\del q}{\del x_0}\{x,z\}_{(z_0,\eta)} + \dfrac{\del q}{\del z_0}\{x,z\}_{(\eta,x_0)} \rbr \right)\dl y
\end{align}
Since $q$ is of the form $q=2\pi+ q^{(2)}\delta^2 +O(\delta^4)$ \eqref{qdef},
\begin{align}
    \{x,z\}_{(x_0,z_0)}=& \, 1 + \delta \eta \lbr \dfrac{\del x_1}{\del x_0} +\dfrac{\del z_1}{\del z_0}\rbr +O(\delta^2),\nonumber \\
    \{x,z\}_{(z_0,\eta)}=& -\delta x_1 +O(\delta^2), \quad
    \{x,z\}_{(\eta,x_0)}= -\delta z_1 +O(\delta^2), \\
     \{q,z\}_{(x_0,z_0)}=& \: \delta^2 \dfrac{\del q^{(2)}}{\del x_0} + +O(\delta^3), \quad \{x,q\}_{(x_0,z_0)}= \: \delta^2 \dfrac{\del q^{(2)}}{\del z_0} +O(\delta^3). \nonumber
\end{align}
Therefore, 
\begin{align}
    \dl (q/\delta^2) &= \lbr \dfrac{\del q^{(2)}}{\del x_0}\dl x + \dfrac{\del q^{(2)}}{\del z_0}\dl z\rbr +O(\delta) \nonumber \\
    \B \times \dl B \cdot \dl (q/\delta^2) &=\delta \{q^{(2)},B_1\}_{(x_0,z_0)} + O(\delta)
\end{align}
From \eqref{B1_def}, \eqref{qdef} we get $\delta B_1=\sqrt{q^{(2)}}$. Therefore, 
\begin{align}
    \B \times \dl B \cdot \dl (q/\delta^2)=O(\delta).
\end{align}

\bibliographystyle{jpp}

\bibliography{plasmalit} 

\begin{thebibliography}{69}
\expandafter\ifx\csname natexlab\endcsname\relax\def\natexlab#1{#1}\fi
\def\au#1{#1} \def\ed#1{#1} \def\yr#1{#1}\def\at#1{#1}\def\jt#1{\textit{#1}}
  \def\bt#1{#1}\def\bvol#1{\textbf{#1}} \def\vol#1{#1} \def\pg#1{#1}
  \def\publ#1{#1}\def\arxiv#1{#1}\def\org#1{#1}\def\st#1{\textit{#1}}

\bibitem[Akerson {\em et~al.\/}(2016)Akerson, Bader, Hegna, Schmitz, Stephey,
  Anderson, Anderson \& Likin]{HSX_akerson2016three}
{\sc \au{Akerson, AR}, \au{Bader, A}, \au{Hegna, CC}, \au{Schmitz, O},
  \au{Stephey, LA}, \au{Anderson, DT}, \au{Anderson, FSB} \& \au{Likin, KM}}
  \yr{2016}  \at{Three-dimensional scrape off layer transport in the helically
  symmetric experiment hsx}.  \jt{Plasma Physics and Controlled Fusion}
  \bvol{58}~(8),  \pg{084002}.

\bibitem[Andreeva(2002)]{andreevaW7Xvacuum}
{\sc \au{Andreeva, T}} \yr{2002}  \bt{Vacuum magnetic configurations of
  wendelstein 7-x}. {\em Tech. Rep.\/}.  \org{Max-Planck-Institut fuer
  Plasmaphysik}.

\bibitem[Bader {\em et~al.\/}(2019)Bader, Drevlak, Anderson, Faber, Hegna,
  Likin, Schmitt \& Talmadge]{bader2019stellarator}
{\sc \au{Bader, Aaron}, \au{Drevlak, M}, \au{Anderson, DT}, \au{Faber, BJ},
  \au{Hegna, CC}, \au{Likin, KM}, \au{Schmitt, JC} \& \au{Talmadge, JN}}
  \yr{2019}  \at{Stellarator equilibria with reactor relevant energetic
  particle losses}.  \jt{Journal of Plasma Physics}  \bvol{85}~(5).

\bibitem[Bauer {\em et~al.\/}(2012)Bauer, Betancourt \&
  Garabedian]{Garabedian2012computational}
{\sc \au{Bauer, Frances}, \au{Betancourt, Octavio} \& \au{Garabedian, Paul}}
  \yr{2012} {\em A computational method in plasma physics\/}.  \publ{Springer
  Science \& Business Media}.

\bibitem[Beidler \& Maa{\ss}berg(2001)]{beidler2001improved}
{\sc \au{Beidler, CD} \& \au{Maa{\ss}berg, H}} \yr{2001}  \at{An improved
  formulation of the ripple-averaged kinetic theory of neoclassical transport
  in stellarators}.  \jt{Plasma physics and controlled fusion}  \bvol{43}~(8),
  \pg{1131}.

\bibitem[BenDaniel(1965)]{benDaniel1965nonexistence}
{\sc \au{BenDaniel, DJ}} \yr{1965}  \at{Nonexistence of isotropic $\oint
  dl/{B}$ equilibria}.  \jt{The Physics of Fluids}  \bvol{8}~(8),
  \pg{1567--1568}.

\bibitem[Boozer(1995)]{boozer1995quasi}
{\sc \au{Boozer, Allen~H}} \yr{1995}  \at{Quasi-helical symmetry in
  stellarators}.  \jt{Plasma Physics and Controlled Fusion}  \bvol{37}~(11A),
  \pg{A103}.

\bibitem[Brakel {\em et~al.\/}(1997)Brakel, Anton, Baldzuhn, Burhenn, Erckmann,
  Fiedler, Geiger, Hartfuss, Heinrich, Hirsch, Jaenicke, Kick, K\"uhner,
  Maa{\ss}berg, Stroth, Wagner, Weller, {W7-AS Team}, {ECRH Group} \&
  {NBI-Group}]{brakel1997W7AS_EB_shear}
{\sc \au{Brakel, R}, \au{Anton, M}, \au{Baldzuhn, J}, \au{Burhenn, R},
  \au{Erckmann, V}, \au{Fiedler, S}, \au{Geiger, J}, \au{Hartfuss, HJ},
  \au{Heinrich, O}, \au{Hirsch, M}, \au{Jaenicke, R}, \au{Kick, M},
  \au{K\"uhner, G}, \au{Maa{\ss}berg, H}, \au{Stroth, U}, \au{Wagner, F},
  \au{Weller, A}, \au{{W7-AS Team}}, \au{{ECRH Group}} \& \au{{NBI-Group}}}
  \yr{1997}  \at{Confinement in {W7-AS} and the role of radial electric field
  and magnetic shear}.  \jt{Plasma Physics and Controlled Fusion}
  \bvol{39}~(12B),  \pg{B273}.

\bibitem[Brakel \& the
  {W7-AS}~Team(2002)]{brakel2002energytransp_rational_iota_W7AS}
{\sc \au{Brakel, R} \& \au{the {W7-AS}~Team}} \yr{2002}  \at{Electron energy
  transport in the presence of rational surfaces in the {Wendelstein 7-AS}
  stellarator}.  \jt{Nuclear fusion}  \bvol{42}~(7),  \pg{903}.

\bibitem[Brizard(2011)]{brizard2011action_pendulum}
{\sc \au{Brizard, Alain~J}} \yr{2011}  \at{Action-angle coordinates for the
  pendulum problem}.  \jt{arXiv preprint arXiv:1108.4970} .

\bibitem[Burby {\em et~al.\/}(2019)Burby, Kallinikos \&
  MacKay]{burby_Kallinikos_MACKAY2019_Math_QS}
{\sc \au{Burby, Joshua~W}, \au{Kallinikos, Nikos} \& \au{MacKay, Robert~S}}
  \yr{2019}  \at{Some mathematics for quasi-symmetry}.  \jt{arXiv preprint
  arXiv:1912.06468} .

\bibitem[Canik {\em et~al.\/}(2007)Canik, Anderson, Anderson, Likin, Talmadge
  \& Zhai]{HSX_canik2007exp_neoclassical}
{\sc \au{Canik, JM}, \au{Anderson, DT}, \au{Anderson, FSB}, \au{Likin, KM},
  \au{Talmadge, JN} \& \au{Zhai, K}} \yr{2007}  \at{Experimental demonstration
  of improved neoclassical transport with quasihelical symmetry}.  \jt{Physical
  review letters}  \bvol{98}~(8),  \pg{085002}.

\bibitem[Cary(1982)]{cary1982vacuum}
{\sc \au{Cary, John~R}} \yr{1982}  \at{Vacuum magnetic fields with dense flux
  surfaces}.  \jt{Physical Review Letters}  \bvol{49}~(4),  \pg{276}.

\bibitem[Cary {\em et~al.\/}(1988)Cary, Hedrick \&
  Tolliver]{cary_Hedrick_1988orbits}
{\sc \au{Cary, John~R}, \au{Hedrick, CL} \& \au{Tolliver, JS}} \yr{1988}
  \at{Orbits in asymmetric toroidal magnetic fields}.  \jt{The Physics of
  fluids}  \bvol{31}~(6),  \pg{1586--1600}.

\bibitem[Cary \& Littlejohn(1983)]{caryLittlejohn1983Hamiltonian_B}
{\sc \au{Cary, John~R} \& \au{Littlejohn, Robert~G}} \yr{1983}
  \at{Noncanonical hamiltonian mechanics and its application to magnetic field
  line flow}.  \jt{Annals of Physics}  \bvol{151}~(1),  \pg{1--34}.

\bibitem[Cary \& Shasharina(1997{\natexlab{{\em
  a\/}}})]{caryshashrinaPRL1997helical}
{\sc \au{Cary, John~R} \& \au{Shasharina, Svetlana~G}} \yr{1997{\natexlab{{\em
  a\/}}}}  \at{Helical plasma confinement devices with good confinement
  properties}.  \jt{Physical review letters}  \bvol{78}~(4),  \pg{674}.

\bibitem[Cary \& Shasharina(1997{\natexlab{{\em
  b\/}}})]{caryshasharina1997omnigenity}
{\sc \au{Cary, John~R} \& \au{Shasharina, Svetlana~G}} \yr{1997{\natexlab{{\em
  b\/}}}}  \at{Omnigenity and quasihelicity in helical plasma confinement
  systems}.  \jt{Physics of Plasmas}  \bvol{4}~(9),  \pg{3323--3333}.

\bibitem[Feng {\em et~al.\/}(2011)Feng, Kobayashi, Lunt \&
  Reiter]{feng2011comparison}
{\sc \au{Feng, Y}, \au{Kobayashi, M}, \au{Lunt, T} \& \au{Reiter, D}} \yr{2011}
   \at{Comparison between stellarator and tokamak divertor transport}.
  \jt{Plasma physics and controlled fusion}  \bvol{53}~(2),  \pg{024009}.

\bibitem[Feng {\em et~al.\/}(2006)Feng, Sardei, Grigull, McCormick, Kisslinger
  \& Reiter]{feng2006physics}
{\sc \au{Feng, Y}, \au{Sardei, F}, \au{Grigull, P}, \au{McCormick, K},
  \au{Kisslinger, J} \& \au{Reiter, D}} \yr{2006}  \at{Physics of island
  divertors as highlighted by the example of w7-as}.  \jt{Nuclear fusion}
  \bvol{46}~(8),  \pg{807}.

\bibitem[Freidberg(1982)]{freidberg_idealMHD}
{\sc \au{Freidberg, Jeffrey~P}} \yr{1982}  \at{Ideal magnetohydrodynamic theory
  of magnetic fusion systems}.  \jt{Reviews of Modern Physics}  \bvol{54}~(3),
  \pg{801}.

\bibitem[Gardner(1959)]{gardner1959adiabatic}
{\sc \au{Gardner, CS}} \yr{1959}  \at{Adiabatic invariants of periodic
  classical systems}.  \jt{Physical Review}  \bvol{115}~(4),  \pg{791}.

\bibitem[Garren \& Boozer(1991)]{garrenBoozer1991existence}
{\sc \au{Garren, DA} \& \au{Boozer, Allen~H}} \yr{1991}  \at{Existence of
  quasihelically symmetric stellarators}.  \jt{Physics of Fluids B: Plasma
  Physics}  \bvol{3}~(10),  \pg{2822--2834}.

\bibitem[Gori {\em et~al.\/}(1996)Gori, Lotz \&
  N{\"u}hrenberg]{gori_Lotz_varenna1996}
{\sc \au{Gori, S}, \au{Lotz, W} \& \au{N{\"u}hrenberg, J}} \yr{1996} Theory of
  fusion plasmas, int.  \bt{In {\em Workshop, Varenna\/}}.

\bibitem[Grad(1967)]{grad1967toroidal}
{\sc \au{Grad, Harold}} \yr{1967}  \at{Toroidal containment of a plasma}.
  \jt{The Physics of Fluids}  \bvol{10}~(1),  \pg{137--154}.

\bibitem[Grad(1971)]{grad1971plasma}
{\sc \au{Grad, Harold}} \yr{1971} Plasma containment in closed line systems.
  \bt{In {\em Plasma Physics and Controlled Nuclear Fusion Research 1971. Vol.
  III. Proceedings of the Fourth International Conference on Plasma Physics and
  Controlled Nuclear Fusion Research\/}}.

\bibitem[Grad(1973)]{grad1973magnetofluid}
{\sc \au{Grad, Harold}} \yr{1973}  \at{Magnetofluid-dynamic spectrum and low
  shear stability}.  \jt{Proceedings of the National Academy of Sciences}
  \bvol{70}~(12),  \pg{3277--3281}.

\bibitem[Hall \& McNamara(1975)]{hall1975three}
{\sc \au{Hall, Laurence~S} \& \au{McNamara, Brendan}} \yr{1975}
  \at{Three-dimensional equilibrium of the anisotropic, finite-pressure
  guiding-center plasma: Theory of the magnetic plasma}.  \jt{The Physics of
  Fluids}  \bvol{18}~(5),  \pg{552--565}.

\bibitem[Hameiri(1980)]{hameiri1980_MHD_local_stability}
{\sc \au{Hameiri, Eliezer}} \yr{1980}  \at{Local sufficient condition for
  magnetohydrodynamic stability of closed line systems}.  \jt{The Physics of
  Fluids}  \bvol{23}~(5),  \pg{889--894}.

\bibitem[Hastie {\em et~al.\/}(1967)Hastie, Taylor \& Haas]{taylor_hastie}
{\sc \au{Hastie, RJ}, \au{Taylor, JB} \& \au{Haas, FA}} \yr{1967}
  \at{Adiabatic invariants and the equilibrium of magnetically trapped
  particles}.  \jt{Annals of Physics}  \bvol{41}~(2),  \pg{302--338}.

\bibitem[Hazeltine \& Meiss(2003)]{hazeltine_meiss2003plasma_confinement_book}
{\sc \au{Hazeltine, Richard~D} \& \au{Meiss, James~D}} \yr{2003} {\em Plasma
  confinement\/}.  \publ{Courier Corporation}.

\bibitem[Helander(2014)]{helander2014theory}
{\sc \au{Helander, Per}} \yr{2014}  \at{Theory of plasma confinement in
  non-axisymmetric magnetic fields}.  \jt{Reports on Progress in Physics}
  \bvol{77}~(8),  \pg{087001}.

\bibitem[Helander {\em et~al.\/}(2012)Helander, Beidler, Bird, Drevlak, Feng,
  Hatzky, Jenko, Kleiber, Proll, Turkin \&
  Xanthopoulos]{helander2012comparison}
{\sc \au{Helander, Per}, \au{Beidler, CD}, \au{Bird, TM}, \au{Drevlak, M},
  \au{Feng, Y}, \au{Hatzky, R}, \au{Jenko, F}, \au{Kleiber, R}, \au{Proll,
  JHE}, \au{Turkin, Yu} \& \au{Xanthopoulos, P}} \yr{2012}  \at{Stellarator and
  tokamak plasmas: a comparison}.  \jt{Plasma Physics and Controlled Fusion}
  \bvol{54}~(12),  \pg{124009}.

\bibitem[Henneberg {\em et~al.\/}(2019)Henneberg, Drevlak \&
  Helander]{henneberg2019fast_particle}
{\sc \au{Henneberg, SA}, \au{Drevlak, M} \& \au{Helander, P}} \yr{2019}
  \at{Improving fast-particle confinement in quasi-axisymmetric stellarator
  optimization}.  \jt{Plasma Physics and Controlled Fusion}  \bvol{62}~(1),
  \pg{014023}.

\bibitem[Hirsch {\em et~al.\/}(2008)Hirsch, Baldzuhn, Beidler, Brakel, Burhenn,
  Dinklage, Ehmler, Endler, Erckmann, Feng, Geiger, Giannone, Grieger, Grigull,
  Hartfu{\ss}, Hartmann, Jaenicke, K\"onig, Laqua, Maa{\ss}berg, McCormick,
  Sardei, Speth, Stroth, Wagner, Weller, Werner, Wobig, Zoletnik \& the {W7-AS
  Team}]{hirsch2008majorW7AS}
{\sc \au{Hirsch, M}, \au{Baldzuhn, J}, \au{Beidler, C}, \au{Brakel, R},
  \au{Burhenn, R}, \au{Dinklage, A}, \au{Ehmler, H}, \au{Endler, M},
  \au{Erckmann, V}, \au{Feng, Y}, \au{Geiger, J}, \au{Giannone, L},
  \au{Grieger, G}, \au{Grigull, P}, \au{Hartfu{\ss}, H-J}, \au{Hartmann, D},
  \au{Jaenicke, R}, \au{K\"onig, R}, \au{Laqua, H~P}, \au{Maa{\ss}berg, H},
  \au{McCormick, K}, \au{Sardei, F}, \au{Speth, E}, \au{Stroth, U}, \au{Wagner,
  F}, \au{Weller, A}, \au{Werner, A}, \au{Wobig, H}, \au{Zoletnik, S} \&
  \au{the {W7-AS Team}}} \yr{2008}  \at{Major results from the stellarator
  wendelstein 7-as}.  \jt{Plasma Physics and Controlled Fusion}  \bvol{50}~(5).

\bibitem[Hudson \& Kraus(2017)]{hudsonKrauss20173D_cont_B}
{\sc \au{Hudson, SR} \& \au{Kraus, BF}} \yr{2017}  \at{Three-dimensional
  magnetohydrodynamic equilibria with continuous magnetic fields}.  \jt{Journal
  of Plasma Physics}  \bvol{83}~(4).

\bibitem[Imbert-Gerard {\em et~al.\/}(2019)Imbert-Gerard, Paul \&
  Wright]{Lima_Paul_Wright_2019introduction}
{\sc \au{Imbert-Gerard, Lise-Marie}, \au{Paul, Elizabeth} \& \au{Wright,
  Adelle}} \yr{2019}  \at{An introduction to symmetries in stellarators}.
  \jt{arXiv preprint arXiv:1908.05360} .

\bibitem[Jaquiery \& Sengupta(2019)]{sengupta_erin2018}
{\sc \au{Jaquiery, Erin} \& \au{Sengupta, Wrick}} \yr{2019}  \at{Low-shear
  three-dimensional equilibria in a periodic cylinder}.  \jt{Journal of Plasma
  Physics}  \bvol{85}~(1),  \pg{905850115}.

\bibitem[Kim {\em et~al.\/}(2019)Kim, McFadden \& Cerfon]{kim2019elimination}
{\sc \au{Kim, Eugenia}, \au{McFadden, Geoffrey} \& \au{Cerfon, Antoine}}
  \yr{2019}  \at{Elimination of mhd current sheets by modifications to the
  plasma wall in a fixed boundary model}.  \jt{arXiv preprint arXiv:1911.01856}
  .

\bibitem[Klinger {\em et~al.\/}(2016)Klinger, Alonso, Bozhenkov, Burhenn,
  Dinklage, Fuchert, Geiger, Grulke, Langenberg, Hirsch {\em
  et~al.\/}]{w7x_klinger2016performance}
{\sc \au{Klinger, Thomas}, \au{Alonso, A}, \au{Bozhenkov, S}, \au{Burhenn, R},
  \au{Dinklage, A}, \au{Fuchert, G}, \au{Geiger, J}, \au{Grulke, O},
  \au{Langenberg, A}, \au{Hirsch, M} \& \au{others}} \yr{2016}  \at{Performance
  and properties of the first plasmas of wendelstein 7-x}.  \jt{Plasma Physics
  and Controlled Fusion}  \bvol{59}~(1),  \pg{014018}.

\bibitem[Klinger {\em et~al.\/}(2019)Klinger, Andreeva, Bozhenkov, Brandt,
  Burhenn, Buttensch{\"o}n, Fuchert, Geiger, Grulke, Laqua {\em
  et~al.\/}]{klinger2019overview}
{\sc \au{Klinger, T}, \au{Andreeva, T}, \au{Bozhenkov, S}, \au{Brandt, C},
  \au{Burhenn, R}, \au{Buttensch{\"o}n, B}, \au{Fuchert, G}, \au{Geiger, B},
  \au{Grulke, O}, \au{Laqua, HP} \& \au{others}} \yr{2019}  \at{Overview of
  first wendelstein 7-x high-performance operation}.  \jt{Nuclear Fusion}
  \bvol{59}~(11),  \pg{112004}.

\bibitem[K{\"o}nig {\em et~al.\/}(2002)K{\"o}nig, Grigull, McCormick, Feng,
  Kisslinger, Komori, Masuzaki, Matsuoka, Obiki, Ohyabu {\em
  et~al.\/}]{konig2002divertor}
{\sc \au{K{\"o}nig, R}, \au{Grigull, P}, \au{McCormick, K}, \au{Feng, Y},
  \au{Kisslinger, J}, \au{Komori, A}, \au{Masuzaki, Suguru}, \au{Matsuoka, K},
  \au{Obiki, T}, \au{Ohyabu, N} \& \au{others}} \yr{2002}  \at{The divertor
  program in stellarators}.  \jt{Plasma physics and controlled fusion}
  \bvol{44}~(11),  \pg{2365}.

\bibitem[Kruskal \& Kulsrud(1958)]{kruskal_Kulsrud_1958equilibrium}
{\sc \au{Kruskal, Martin~David} \& \au{Kulsrud, RM}} \yr{1958}  \at{Equilibrium
  of a magnetically confined plasma in a toroid}.  \jt{The Physics of Fluids}
  \bvol{1}~(4),  \pg{265--274}.

\bibitem[Landreman(2019)]{Landreman_Simons_Summer_lecture}
{\sc \au{Landreman, Matt}} \yr{2019}  \bt{Quasisymmetry: A hidden symmetry of
  magnetic fields}. Simons Collaboration on Hidden Symmetries and Fusion
  Energy,  \publ{hiddensymmetries.princeton.edu},
  https://hiddensymmetries.princeton.edu/summer-school/lecture-notes.

\bibitem[Landreman \& Catto(2012)]{landremancatto2012omnigenity}
{\sc \au{Landreman, Matt} \& \au{Catto, Peter~J}} \yr{2012}  \at{Omnigenity as
  generalized quasisymmetry a}.  \jt{Physics of Plasmas}  \bvol{19}~(5),
  \pg{056103}.

\bibitem[Landreman \& Sengupta(2018)]{landreman_Sengupta2018direct}
{\sc \au{Landreman, Matt} \& \au{Sengupta, Wrick}} \yr{2018}  \at{Direct
  construction of optimized stellarator shapes. part 1. theory in cylindrical
  coordinates}.  \jt{Journal of Plasma Physics}  \bvol{84}~(6).

\bibitem[Landreman \& Sengupta(2019)]{landreman_Sengupta2019_2nd_order}
{\sc \au{Landreman, Matt} \& \au{Sengupta, Wrick}} \yr{2019}  \at{Constructing
  stellarators with quasisymmetry to high order}.  \jt{Journal of Plasma
  Physics}  \bvol{85}~(6).

\bibitem[Landreman {\em et~al.\/}(2019)Landreman, Sengupta \&
  Plunk]{landreman_Sengupta_Plunck2019direct}
{\sc \au{Landreman, Matt}, \au{Sengupta, Wrick} \& \au{Plunk, Gabriel~G}}
  \yr{2019}  \at{Direct construction of optimized stellarator shapes. part 2.
  numerical quasisymmetric solutions}.  \jt{Journal of Plasma Physics}
  \bvol{85}~(1).

\bibitem[Lortz(1970)]{lortz1970existenz}
{\sc \au{Lortz, Dietrich}} \yr{1970}  \at{{\"U}ber die existenz toroidaler
  magnetohydrostatischer gleichgewichte ohne rotationstransformation}.
  \jt{Zeitschrift f{\"u}r angewandte Mathematik und Physik ZAMP}
  \bvol{21}~(2),  \pg{196--211}.

\bibitem[Mynick(2006)]{mynick2006transport}
{\sc \au{Mynick, HE}} \yr{2006}  \at{Transport optimization in stellarators}.
  \jt{Physics of plasmas}  \bvol{13}~(5),  \pg{058102}.

\bibitem[Nelson \& Spies(1974)]{nelson_spies_1974_suff_MHD_closed}
{\sc \au{Nelson, DB} \& \au{Spies, GO}} \yr{1974}  \at{Improved sufficient
  conditions for magnetohydrodynamic stability with closed field lines}.
  \jt{The Physics of Fluids}  \bvol{17}~(11),  \pg{2133--2134}.

\bibitem[Newcomb(1959)]{newcomb1959magnetic}
{\sc \au{Newcomb, William~A}} \yr{1959}  \at{Magnetic differential equations}.
  \jt{The Physics of Fluids}  \bvol{2}~(4),  \pg{362--365}.

\bibitem[N{\"u}hrenberg(2010)]{nuhrenberg2010isodynamic}
{\sc \au{N{\"u}hrenberg, J{\"u}rgen}} \yr{2010}  \at{Development of
  quasi-isodynamic stellarators}.  \jt{Plasma Physics and Controlled Fusion}
  \bvol{52}~(12),  \pg{124003}.

\bibitem[N{\"u}hrenberg {\em et~al.\/}(1994)N{\"u}hrenberg, Sindoni, Lotz,
  Troyon, Gori \& Vaclavik]{nuhrenberg1994quasiaxi}
{\sc \au{N{\"u}hrenberg, J}, \au{Sindoni, E}, \au{Lotz, W}, \au{Troyon, F},
  \au{Gori, S} \& \au{Vaclavik, J}} \yr{1994} Quasi-axisymmetric tokamaks.
  \bt{In {\em Proc. of the Joint Varenna-Lausanne International Workshop on
  Theory of Fusion Plasmas\/}},  \pg{pp. 3--12}. Ed. Compositori.

\bibitem[N{\"u}hrenberg \& Zille(1988)]{nuhrenberg1988quasihelical}
{\sc \au{N{\"u}hrenberg, J} \& \au{Zille, R}} \yr{1988}  \at{Quasi-helically
  symmetric toroidal stellarators}.  \jt{Physics Letters A}  \bvol{129}~(2),
  \pg{113--117}.

\bibitem[Okamura {\em et~al.\/}(2001)Okamura, Matsuoka, Nishimura, Isobe,
  Nomura, Suzuki, Shimizu, Murakami, Nakajima, Yokoyama, Fujisawa, Ida, Itoh,
  Merkel, Drevlak, Zille, Gori \& N\"{u}hrenberg]{okamura2001quasiaxi}
{\sc \au{Okamura, S}, \au{Matsuoka, K}, \au{Nishimura, S}, \au{Isobe, M},
  \au{Nomura, I}, \au{Suzuki, C}, \au{Shimizu, A}, \au{Murakami, S},
  \au{Nakajima, N}, \au{Yokoyama, M}, \au{Fujisawa, A}, \au{Ida, K}, \au{Itoh,
  K}, \au{Merkel, P}, \au{Drevlak, M}, \au{Zille, R}, \au{Gori, S} \&
  \au{N\"{u}hrenberg, J}} \yr{2001}  \at{Physics and engineering design of the
  low aspect ratio quasi-axisymmetric stellarator chs-qa}.  \jt{Nuclear fusion}
   \bvol{41}~(12),  \pg{1865}.

\bibitem[Plunk \& Helander(2018)]{plunk2018quasiaxi_vacuum}
{\sc \au{Plunk, GG} \& \au{Helander, Per}} \yr{2018}  \at{Quasi-axisymmetric
  magnetic fields: weakly non-axisymmetric case in a vacuum}.  \jt{Journal of
  Plasma Physics}  \bvol{84}~(2).

\bibitem[Plunk {\em et~al.\/}(2019)Plunk, Landreman \&
  Helander]{plunk_Landreman_2019near_axis3}
{\sc \au{Plunk, Gabriel~G}, \au{Landreman, Matt} \& \au{Helander, Per}}
  \yr{2019}  \at{Direct construction of optimized stellarator shapes. part 3.
  omnigenity near the magnetic axis}.  \jt{Journal of Plasma Physics}
  \bvol{85}~(6).

\bibitem[Sengupta \& Weitzner(2018)]{sengupta_weitzner2018}
{\sc \au{Sengupta, Wrick} \& \au{Weitzner, Harold}} \yr{2018}  \at{Radial
  confinement of deeply trapped particles in a non-symmetric
  magnetohydrodynamic equilibrium}.  \jt{Physics of Plasmas}  \bvol{25}~(2),
  \pg{022506}.

\bibitem[Sengupta \& Weitzner(2019)]{sengupta_weitzner2019_lowshear_vacuum}
{\sc \au{Sengupta, Wrick} \& \au{Weitzner, Harold}} \yr{2019}  \at{Low-shear
  three-dimensional equilibria and vacuum magnetic fields with flux surfaces}.
  \jt{Journal of Plasma Physics}  \bvol{85}~(2).

\bibitem[Skovoroda(2005)]{skovoroda20053dimproved}
{\sc \au{Skovoroda, AA}} \yr{2005}  \at{3d toroidal geometry of currentless
  magnetic configurations with improved confinement}.  \jt{Plasma physics and
  controlled fusion}  \bvol{47}~(11),  \pg{1911}.

\bibitem[Spies(1974)]{spies1974nonlinear_stability_closed}
{\sc \au{Spies, GO}} \yr{1974}  \at{Nonlinear magnetohydrodynamic stability and
  plasma containment with closed field lines}.  \jt{The Physics of Fluids}
  \bvol{17}~(6),  \pg{1188--1197}.

\bibitem[Spies(1979)]{spies1979low_report}
{\sc \au{Spies, G{\"u}nther~O}} \yr{1979}  \at{Low-shear magnetohydrodynamic
  stability.}  \jt{Report 6/185} .

\bibitem[Subbotin {\em et~al.\/}(2006)Subbotin, Mikhailov, Shafranov, Isaev,
  N{\"u}hrenberg, N{\"u}hrenberg, Zille, Nemov, Kasilov, Kalyuzhnyj {\em
  et~al.\/}]{subbotin2006integrated}
{\sc \au{Subbotin, AA}, \au{Mikhailov, MI}, \au{Shafranov, VD}, \au{Isaev,
  M~Yu}, \au{N{\"u}hrenberg, C}, \au{N{\"u}hrenberg, J}, \au{Zille, R},
  \au{Nemov, VV}, \au{Kasilov, SV}, \au{Kalyuzhnyj, VN} \& \au{others}}
  \yr{2006}  \at{Integrated physics optimization of a quasi-isodynamic
  stellarator with poloidally closed contours of the magnetic field strength}.
  \jt{Nuclear fusion}  \bvol{46}~(11),  \pg{921}.

\bibitem[Taylor(1963)]{taylor1963_stability_cusp}
{\sc \au{Taylor, JB}} \yr{1963}  \at{Some stable plasma equilibria in combined
  mirror-cusp fields}.  \jt{The Physics of Fluids}  \bvol{6}~(11),
  \pg{1529--1536}.

\bibitem[Taylor \& Hastie(1968)]{taylor_hastie_1968stability}
{\sc \au{Taylor, JB} \& \au{Hastie, RJ}} \yr{1968}  \at{Stability of general
  plasma equilibria-i formal theory}.  \jt{Plasma Physics}  \bvol{10}~(5),
  \pg{479}.

\bibitem[Weitzner(2014)]{weitzner2014ideal}
{\sc \au{Weitzner, Harold}} \yr{2014}  \at{Ideal magnetohydrodynamic
  equilibrium in a non-symmetric topological torus}.  \jt{Physics of Plasmas}
  \bvol{21}~(2),  \pg{022515}.

\bibitem[Weitzner(2016)]{weitzner2016expansions}
{\sc \au{Weitzner, Harold}} \yr{2016}  \at{Expansions of non-symmetric toroidal
  magnetohydrodynamic equilibria}.  \jt{Physics of Plasmas}  \bvol{23}~(6),
  \pg{062512}.

\bibitem[Weitzner \& Sengupta(2019)]{weitzner_sengupta2019exact}
{\sc \au{Weitzner, Harold} \& \au{Sengupta, Wrick}} \yr{2019}  \at{Exact
  non-symmetric closed line vacuum magnetic fields in a topological torus}.
  \jt{arXiv preprint arXiv:1909.01890 (accepted for publication in Physics of
  Plasmas)} .

\bibitem[Wobig(1987)]{wobig1987localized_pert_w7A}
{\sc \au{Wobig, Horst}} \yr{1987}  \at{Magnetic surfaces and localized
  perturbations in the {Wendelstein VII-A} stellarator}.  \jt{Zeitschrift
  f{\"u}r Naturforschung A}  \bvol{42}~(10),  \pg{1054--1066}.

\end{thebibliography}

\end{document}